\documentclass[a4paper,10pt]{article}

\usepackage{graphicx}
\usepackage{subeqn}
\usepackage{amssymb}
\usepackage{lineno}
\usepackage{bm}
\usepackage{array}
\usepackage{color}
\usepackage{subfigure}

\date{ }

\begin{document}

\title{On the Use of Fractional Calculus for the Probabilistic Characterization of Random Variables
\footnote{Publication info: Cottone G., Di Paola M.,
On the use of fractional calculus for the probabilistic characterization of random variables,
Probabilistic Engineering Mechanics, 
Volume 24, Issue 3, 2009,
Pages 321-330, ISSN 0266-8920, 
10.1016/j.probengmech.2008.08.002.} }

\author{Giulio Cottone \footnote{E-mail: giulio.cottone@tum.de; giuliocottone@yahoo.it} and Mario Di Paola \\
\small Dipartimento di Ingegneria Civile, Aerospaziale ed Ambientale, \\ 
\small Universit\'{a} degli Studi di Palermo, Viale delle Scienze, 90128 Palermo, Italy\\
}


\maketitle


\begin{abstract}
In this paper, the classical problem of the probabilistic characterization of a random variable is re-examined. A random variable is usually described by the probability density function (PDF) or by its Fourier transform, namely the characteristic function (CF). The CF can be further expressed by a Taylor series involving the moments of the random variable. However, in some circumstances, the moments do not exist and the Taylor expansion of the CF is useless. This happens for example in the case of $\alpha$--stable random variables.
Here, the problem of representing the CF or the PDF of random variables (r.vs) is examined by introducing fractional calculus. Two very remarkable results are obtained. Firstly, it is shown that the fractional derivatives of the CF in zero coincide with fractional moments. This is true also in case of CF not derivable in zero (like the CF of  $\alpha$--stable r.vs). Moreover, it is shown that the CF may be represented by a generalized Taylor expansion involving fractional moments. The generalized Taylor series proposed is also able to represent the PDF in a perfect dual representation to that in terms of CF. The PDF representation in terms of fractional moments is especially accurate in the tails and this is very important in engineering problems, like estimating structural safety.
\end{abstract}

\section{Introduction}

\noindent In many cases of engineering interest, it is useful to describe strength or mechanical properties of materials, geometrical features of structures or loads and so on, as random variables. A random variable is fully characterized by the PDF or by its spectral counterpart, namely the CF, or by the moments (or cumulants) of every order. Moments and cumulants are related to the coefficients of the CityplaceTaylor expansion in zero of the CF and of the log CF, respectively.

Yet, the moment representation of the CF is not always feasible. Indeed, if the derivatives in zero of the CF do not exist, the Taylor moments series is meaningless. For example, the CF of $\alpha$--stable random variable [1], [2], is not derivable and consequently such representation does not exist. Another example is the response to non--linear structures under parametric stochastic input, which may have divergent moments starting from a certain order, even if the system is stable in probability. With such moment structure, the CF cannot be restored. From these observations, it might be concluded that moment resolution strategies are often infeasible. 

\noindent In this paper, we investigate on the particular class of moments with complex exponent and we show how powerful this extended class is. To the author's knowledge, such moments have not been investigated in literature. Then, definitions and existence conditions of such moments will be given in the following. Of course, as the PDF is a real function, it follows straightforwardly that if the moments up to a given integer order $n$ exists then all the complex moments with real part less than $n$ and greater than zero must exist. The question is: are such moments useful in reconstructing the CF and the PDF? 

\noindent In order to answer to this question fractional calculus will be adopted. The latter has received a growing interest in the last century and applications in physics and biophysics, in quantum mechanics, in the study of porous systems (gathered in the book [3]), in fracture mechanics [4], in non--local elasticity [5], [6], to cite just few, are available in literature. In stochastic dynamics, the PDF of response to differential equations driven by L\'evy $\alpha$--stable white noise processes is ruled by a fractional differential equation, involving fractional derivative in the diffusive term [7]. Fractional derivatives are encountered also in random vibration with frequency dependent parameter [8] and in the analysis of linear or non--linear systems driven by fractional Brownian motion [9], [10].

\noindent By means of the fractional calculus, we show that the fractional derivatives of the CF in zero coincide with fractional moments, also in the case of CF not derivable in zero (like the CF of $\alpha$--stable r.vs). To the authors' knowledge, the only author giving a relation between fractional moments and fractional derivatives of the CF in zero is Wolfe [11] by using fractional derivatives. However, the expression obtained by Wolfe is given in an integral form quite different from the classical expression relating moments and derivatives in zero of the CF. 

\noindent Indicating with ${\bf {\mathbb C}}$ the set of complex numbers, we will show that fractional derivatives and integrals of order $\gamma \in {\bf {\mathbb C}}$ of the CF calculated in zero are nothing more than particular moments of order $\gamma $ of the random variable. Once this remarkable result is achieved, the usefulness of fractional moments for the probabilistic characterization of random variables is pointed out by using a generalized Taylor integral theorem proposed by Samko et al. [12], which involves fractional derivatives. It is shown that, by means of a suitable discretization, the series here proposed involving moments of complex order looks like the classical Taylor series. A satisfactory representation of the CF is made possible also for the cases in which the CF exhibits a slope discontinuity in zero as in the case of $\alpha$--stable random variables. Moreover, it is well known that, when the classical Taylor series is truncated, the CF exhibits unsatisfactory trends on tails, whilst in this paper it is shown that the fractional Taylor series always fulfils the desired fundamental property that the CF vanishes at infinity. By some easy algebra we will show a very useful dual representation of the PDF by means of complex moments. Finally, we will show by numerical examples that with a finite set of complex moments one can have very good approximations both of the PDF and of the CF. In particular, the PDF is very good approximated in the tails and this is important in problems like the estimation of structural safety.

\section{Preliminary Concepts and Definitions}

\noindent In this section, some well-known concepts on the probabilistic characterization of random variables, as well as some definition of fractional differential calculus, are briefly summarized for clarity's sake and with the aim to introduce appropriate symbols. 

Let $X\in {\bf {\mathbb R}}$ be a real random variable whose probabilistic characterization may be given both by the PDF $p_{X} \left(x\right)$ and by its Fourier transform, namely the CF $\phi _{X} \left(\vartheta \right)$, that is

\begin{equation} 
\label{eq1} 
\phi _{X} \left(\vartheta \right)=E\left[\exp \left(i\vartheta X\right)\right]=\int _{-\infty }^{\infty }\exp \left(i\vartheta x\right)p_{X} \left(x\right)dx  
\end{equation} 
where $\vartheta \in {\mathbb R}$, $i=\sqrt{-1}$ is the imaginary unit and $E\left[ \cdot \right]$ indicates average. Provided that moments $E\left[X^{j} \right]$ with $j=1,2,...$, defined as 
\begin{equation} 
\label{eq2} 
E\left[X^{j} \right]=\int _{-\infty }^{\infty }p_{X} \left(x\right)x^{j} dx  
\end{equation} 
exist, then $\phi _{X} \left(\vartheta \right)$ can be expanded in Taylor series
\begin{equation} 
\label{eq3} 
\phi _{X} \left(\vartheta \right)=\sum _{j=0}^{\infty }\frac{\left(i_{} \vartheta \right)^{j} }{j!} E\left[X^{j} \right]  
\end{equation} 
due to the property
\begin{equation} 
\label{eq4} 
E\left[(iX)^{j} \right]=\left. \frac{d^{j} \phi _{X} \left(\vartheta \right)}{d\vartheta ^{j} } \right|_{\vartheta =0} \,\,\,\,\,\, {\rm{for}} j=0,1,2,... 
\end{equation} 

From equation (\ref{eq4}), it is clear that $E\left[X^{j} \right]$ exists if the j-th derivative in zero of the CF exists. For example, the moments $E\left[\left|X\right|^{p} \right]$ of the $\alpha$-stable random variables do not exist for $p\ge \alpha $ and, since the stability index $\alpha $ ranges from zero up to 2, then the Taylor expansion (\ref{eq3}) cannot be applied. In spite of this, since $E\left[\left|X\right|^{p} \right]$ and $E\left[X^{p} \right]\in {\bf {\mathbb C}}$ with $p<\alpha $ exist, is it possible expanding the CF into some series involving fractional moments of order $p<\alpha $? The answer is affirmative, as we prove after recalling few remarks on fractional calculus. For more exhaustive treatment on fractional calculus, readers are referred to the excellent encyclopaedic book of [12], and to [13], [14]. 
The greatest difficulty one has to overcome dealing with fractional calculus is represented by so many definitions of fractional derivatives present in literature, and each definition has its own peculiarities and technicalities. As it will be stated clearly in the next sections, every definition is good for our purpose. In particular, we recall the definitions of the Riemann-Liouville, Marchaud and Riesz fractional derivatives and integrals since they are useful for the ensuing results.

The Riemann--Liouville (RL) fractional integrals of order $\rho \in {\rm {\mathbb R}\; >0}$, denoted as $\left(I_{\pm }^{\rho } f\right)\left(x\right)$ are defined as follows 
\begin{equation} 
\label{eq5} 
\left(I_{\pm }^{\rho } f\right)\left(x\right)\mathop{=}\limits^{def} \frac{1}{\Gamma \left(\rho \right)} \int _{0}^{\infty }\xi ^{\rho -1} f\left(x\mp \xi \right)d\xi   
\end{equation} 
where $\Gamma \left(\rho \right)\mathop{=}\limits^{def} \int _{0}^{\infty }\xi ^{\rho -1} {\rm exp}\left({\rm -}\xi \right){\rm d}\xi  $ is the Euler gamma function that interpolates the factorial function, that is $\left(n-1\right){\rm !}=\Gamma \left(n\right)$. 
Often, the operators $\left(I_{+}^{\rho } f\right)\left(x\right)$ and $\left(I_{-}^{\rho } f\right)\left(x\right)$ are referred as left and right hand sided, respectively. In some literature, these integrals $\left(I_{\pm }^{\rho } f\right)\left(x\right)$ are also denoted as Liouville-Weyl (LW) fractional integrals. The RL fractional derivatives, denoted as $\left({\mathcal D}_{\pm }^{\rho } f\right)\left(x\right)$, are expressed as
\begin{equation} 
\label{eq6} 
\left({\mathcal D}_{\pm }^{\rho } f\right)\left(x\right)\mathop{=}\limits^{def} \frac{(\pm 1)^{n} }{\Gamma \left(n-\rho \right)} \frac{d^{n} }{dx^{n} } \int _{0}^{\infty }\xi ^{n-\rho -1} f\left(x\mp \xi \right)d\xi   
\end{equation} 
with $\rho \in {\rm {\mathbb R}\; >0}$ and $n=\left[\rho \right]+1$, being $\left[\rho \right]$ the integer part of $\rho $. Examining RL definitions it is clear that the conditions of existence of such operators depend strongly on the behavior of the function $f\left(x\right)$ at $\pm \infty $.

The Marchaud definition of the fractional derivative, denoted as $\left({\bf D}_{\pm }^{\rho } f\right)\left(x\right)$ is expressed as
\begin{equation} 
\label{eq7} 
\left({\bf D}_{\pm }^{\rho } f\right)\left(x\right)=\frac{\left\{\rho \right\}}{\Gamma \left(1-\left\{\rho \right\}\right)} \int _{0}^{\infty }\frac{f^{\left(\left[\rho \right]\right)} \left(x\right)-f^{\left(\left[\rho \right]\right)} \left(x\mp \xi \right)d\xi }{\xi ^{1+\left\{\rho \right\}} }   
\end{equation} 
where $f^{\left(\left[\rho \right]\right)} \left(x\right)$ denotes the derivative of order equal to the integer part of the real number $\rho $, and $\left\{\rho \right\}=\rho -\left[\rho \right]$. It is worth to note that the Marchaud fractional derivative exists also for functions growing at infinity as $\left|x\right|^{\rho -\varepsilon } ,\varepsilon >0$. 

The Riesz fractional integration, denoted as $\left(I_{}^{\rho } f\right)\left(x\right)$ is defined as follows 
\begin{equation} \label{eq8} 
\left(I_{}^{\rho } f\right)\left(x\right)\mathop{=}\limits^{def} \frac{1}{2_{} \Gamma \left(\rho \right)\cos \left(\rho _{} \pi /2\right)} \int _{-\infty }^{\infty }\frac{f\left(\xi \right)d\xi }{\left|\xi -x\right|^{1-\rho } }   
\end{equation} 
with $\rho \in {\rm {\mathbb R}}\, {\rm >}\, {\rm 0}$, $\rho \ne {\rm 1,3,5...}$.The Riesz integral may be expressed by the RL operator as follows
\begin{equation} 
\label{eq9} 
\left(I_{}^{\rho } f\right)\left(x\right)\mathop{=}\limits^{def} \frac{1}{2\cos \left(\rho _{} \pi /2\right)} \left(\left(I_{+}^{\rho } f\right)\left(x\right)+\left(I_{-}^{\rho } f\right)\left(x\right)\right) 
\end{equation} 
and the Riesz fractional derivative denoted as $\left({\mathcal D}^{\rho } f\right)\left(x\right)$ may be represented in terms of Marchaud fractional derivative [12]
\begin{equation} 
\label{eq10} 
\left({\mathcal D}_{}^{\rho } f\right)\left(x\right)=-\frac{1}{2_{} \cos \left(\rho _{} \pi /2\right)} \left(\left({\bf D}_{+}^{\rho } f\right)\left(x\right)+\left({\bf D}_{-}^{\rho } f\right)\left(x\right)\right) 
\end{equation} 

Moreover, the Riesz fractional derivative may be also expressed in terms of RL fractional derivatives as follows [9], [10]
\begin{equation} 
\label{eq11} 
\left({\mathcal D}_{}^{\rho } f\right)\left(x\right)=-\frac{1}{2_{} \cos \left(\rho _{} \pi /2\right)} \left(\left({\mathcal D}_{+}^{\rho } f\right)\left(x\right)+\left({\mathcal D}_{-}^{\rho } f\right)\left(x\right)\right) 
\end{equation} 
The operation of fractional integration, which has been introduced for real $\rho >0$, can be made meaningful also for complex value of the index, say it $\gamma =\rho +i\, \eta $, with $\rho ,\eta \in {\rm {\mathbb R}}$. Indeed, the fractional integrals $\left(I_{\pm }^{\gamma } f\right)\left(x\right)$ and $\left(I_{}^{\gamma } f\right)\left(x\right)$ are properly defined under the condition that $\rho >0$. In the same way, fractional derivatives $\left({\mathcal D}_{\pm }^{\gamma } f\right)\left(x\right)$, $\left({\mathcal D}^{\gamma } f\right)\left(x\right)$ are defined accordingly to relations (\ref{eq6}), (\ref{eq7}), (\ref{eq10}), (\ref{eq11}), having care that $n=\left[Re\gamma \right]=\left[\rho \right]$. It must be clear that integrals or derivatives of complex order $\gamma \in {\rm {\mathbb C}}$ (and $\rho \ne 0$) represent an analytic continuation in the parameter $\gamma $ of fractional integrals and derivatives originally defined for $\eta =Im\gamma =0$. Sufficient conditions on the existence of such operators and extensions to the case of $Re\gamma =0$ are reported in Samko et al.([12], p.38 -- 39). Here, we stress only that the fractional derivatives are derivatives of convolution integral that produces a smoothing on the original function: then, the fractional derivative of a function might exist even if the function is not classically differentiable. 

A very important feature of fractional operators is their behaviour with respect to the Fourier transform. The Fourier transform of a function $f\left(x\right)$ defined in the whole real axis, is a function of the variable $\vartheta $ defined as ${\mathcal F}_{} \left\{f\left(x\right);\vartheta \right\}$ $=\smallint _{-\infty }^{\infty } e^{i_{} x_{} \vartheta } f\left(x\right)dx$, jointly with the inverse transform ${\mathcal F}^{-1} \left\{g\left(\vartheta \right);x\right\}=\left(2_{} \pi \right)^{-1} \smallint _{-\infty }^{\infty } e^{-i_{} x_{} \vartheta } g\left(\vartheta \right)d\vartheta $. Then, it has be proven that the Fourier transform of Riemann--Liouville, Marchaud, and Riesz fractional derivatives of order $\gamma $ are given, respectively, by the relations
\begin{equation} 
\label{eq12} 
{\mathcal F}\left\{\left({\mathcal D}_{\pm }^{\gamma } f\right)\left(x\right);\vartheta \right\}=\left(\mp i_{} \vartheta \right)^{\gamma } {\mathcal F}\left\{f\left(x\right);\vartheta \right\} 
\end{equation} 
\begin{equation} 
\label{eq13} 
{\mathcal F}\left\{\left({\bf D}_{\pm }^{\gamma } f\right)\left(x\right);\vartheta \right\}=\left(\mp i_{} \vartheta \right)^{\gamma } {\mathcal F}\left\{f\left(x\right);\vartheta \right\} 
\end{equation} 
\begin{equation}
\label{eq14} 
{\mathcal F}\left\{\left({\mathcal D}_{}^{\gamma } f\right)\left(x\right);\vartheta \right\}=-\left|\vartheta \right|^{\gamma } {\mathcal F}\left\{f\left(x\right);\vartheta \right\} 
\end{equation} 
where $Re\gamma >0$ (see [12], pp 137; 218).

Analogously, the Fourier transforms of Riemann--Liouville and Riesz fractional integrals are given by the relations
\begin{equation} \label{eq15} 
{\mathcal F}\left\{I_{\pm }^{\gamma } f\left(x\right);\vartheta \right\}=\left(\mp i_{} \vartheta \right)^{-\gamma } {\mathcal F}\left\{f\left(x\right);\vartheta \right\},      0<Re\gamma <1 
\end{equation} 
\begin{equation}
\label{eq16} 
{\rm {\mathcal F}}\left\{I_{}^{\gamma } f\left(x\right);\vartheta \right\}=\left|\vartheta \right|^{-\gamma } {\rm {\mathcal F}}\left\{f\left(x\right);\vartheta \right\},          0<Re\gamma <1 
\end{equation} 

Some of the main properties and rules of the aforementioned fractional derivatives and integrals like Mellin transform and composition rules are reported in Appendix A.

\section{Fractional derivatives and integrals of the characteristic function}
In this section, useful relationships between fractional moments and Riesz fractional integrals and derivatives are derived. In the following of the paper, it will be indicated by $\gamma $ a complex number, $\gamma =\rho +i\eta $, with $\rho $, $\eta $ reals. The function used in the following $\left(\mp i_{} x\right)^{\gamma } $ has to be understood ([12], p.137) as
\begin{equation} 
\label{eq17} 
\left(\mp i_{} x\right)^{\gamma } =\exp \left(\gamma \ln \left|x\right|\mp \frac{\gamma _{} \pi _{} i}{2} sgn\left(x\right)\right),\, \, \, \, \, \, \, \gamma \in {\rm {\mathbb C}},\, Re\gamma >0,\, x\in {\rm {\mathbb R}},        
\end{equation} 
being 
\begin{equation}
\label{eq18} 
sgn\left(x\right)\mathop{=}\limits^{def} \left\{\begin{array}{ccc} {1} & {} & {x>0} \\ {0} & {} & {x=0} \\ {-1} & {} & {x<0} \end{array}\right.  
\end{equation} 
the usual \textit{signum} function of a real variable. Now, following eq.(\ref{eq12}), the Fourier transform of the Riesz fractional derivative of the CF is given as
\begin{equation} 
\label{eq19} 
{\rm {\mathcal F}}\left\{\left({\rm {\mathcal D}}_{\pm }^{\gamma } \phi _{X} \right)\left(\vartheta \right);x\right\}=\left(\mp ix\right)^{\gamma } {\rm {\mathcal F}}\left\{\phi _{X} \left(\vartheta \right);x\right\},           Re\gamma >0 
\end{equation} 
Applying the inverse Fourier transform to both sides of eq.(\ref{eq19}) gives
\begin{equation} 
\label{eq20} 
\left({\rm {\mathcal D}}_{\pm }^{\gamma } \phi _{X} \right)\left(\vartheta \right)={\rm {\mathcal F}}^{-1} \left\{{\rm {\mathcal F}}\left\{\left({\rm {\mathcal D}}_{\pm }^{\gamma } \phi _{X} \right)\left(\vartheta \right);x\right\};\vartheta \right\}=\int _{-\infty }^{\infty }\left(\mp ix\right)^{\gamma } p_{X} \left(x\right)\exp \left(i\vartheta x\right)dx  
\end{equation} 
Finally, by evaluating eq.(\ref{eq20}) in $\vartheta =0$ we obtain the fundamental relation
\begin{equation} 
\label{eq21} 
E\left[\left(\mp iX\right)^{\gamma } \right]=\int _{-\infty }^{\infty }\left(\mp ix\right)^{\gamma } p_{X} \left(x\right)dx =\left({\rm {\mathcal D}}_{\pm }^{\gamma } \phi _{X} \right)\left(0\right);        \gamma \in {\bf {\mathbb C}},Re\gamma >0 
\end{equation} 
stating that the fractional RL derivative of order $\gamma $ of the CF in zero equals the complex moment of order $\gamma $ of the random variable $X$. That is, complex moments are ruled by an expression very similar to that obtained by classical moments (see eq.(\ref{eq4}). 
The easy procedure leading to (\ref{eq21}) may also be used for demonstrating the following identities, starting from the Fourier properties (\ref{eq12})--(\ref{eq16})
\begin{equation} 
\label{eq22} 
\left({\rm {\mathcal D}}_{}^{\gamma } \phi _{X} \right)\left(0\right)=-E\left[\left|X\right|^{\gamma } \right],   Re\gamma >0 
\end{equation} 
\begin{equation} 
\label{eq23} 
\left(I_{}^{\gamma } \phi _{X} \right)\left(0\right)=E\left[\left|X\right|^{-\gamma } \right],   Re\gamma >0 
\end{equation} 
\begin{equation} 
\label{eq24} 
\left({\bf D}_{\pm }^{\gamma } \phi _{X} \right)\left(0\right)=E\left[\left(\mp iX\right)^{\gamma } \right],   Re\gamma >0 
\end{equation} 
\begin{equation} 
\label{eq25} 
\left(I_{\pm }^{\gamma } \phi _{X} \right)\left(0\right)\, =E\left[\left(\mp iX\right)^{-\gamma } \right]  ,   Re\gamma >0 
\end{equation} 
We want to remark that, Wolfe [11] derived fractional moments, by using the Marchaud fractional derivatives in the form
\begin{eqnarray} 
\label{eq26} 
E\left[\left|X\right|^{q} \right]=\frac{i^{-q-1} }{2\pi } Re\left\{\int _{-\infty }^{\infty }\left[\left({\bf D}_{+}^{q} \phi _{X} \right)\left. \left(\vartheta \right)\right|_{\vartheta =\xi } \right. - \left. \left({\bf D}_{+}^{q} \phi _{X} \right)\left. \left(\vartheta \right)\right|_{\vartheta =\xi } \right]\frac{d\xi }{\xi } \right\}+\\ \nonumber
+\, i^{-q} Re\left\{\left({\bf D}_{+}^{q} \phi _{X} \right)\left. \left(\vartheta \right)\right|_{\vartheta =0} \right\} 
\end{eqnarray} 
with $-\infty <q<\infty $. Notice that the structure of the eq.(\ref{eq26}) hides the simplicity of the conceptual connection between moments and characteristic function present in the corresponding eq.(\ref{eq22}). Furthermore, eq.(\ref{eq26}) does not give fractional moments explicitly, because of the integral involved. Eqs.(\ref{eq21})--(\ref{eq25}) have been obtained in the most general context by using complex exponents and can be written for real exponent as well.

It should be noted that moments of the type $E\left[\left(\mp iX\right)^{\pm \gamma } \right]$, after some cumbersome algebra, may be expressed in terms of moments $E\left[X^{\pm \gamma } \right]$, in the form
\begin{equation} 
\label{eq27} 
E\left[\left(iX\right)^{\gamma } \right]=2\cos \left(\frac{\pi \gamma }{2} \right)E\left[\left|X\right|^{\gamma } \right]-i^{-\gamma } E\left[X^{\gamma } \right] 
\end{equation} 
\begin{equation}
\label{eq28} 
E\left[\left(-iX\right)^{\gamma } \right]=i^{-\gamma } E\left[X^{\gamma } \right] 
\end{equation} 
\begin{equation} 
\label{eq29} 
E\left[\left(iX\right)^{-\gamma } \right]=2\cos \left(\frac{\pi \gamma }{2} \right)E\left[\left|X\right|^{-\gamma } \right]-i^{\gamma } E\left[X^{-\gamma } \right] 
\end{equation} 
\begin{equation} 
\label{eq30} 
E\left[\left(-iX\right)^{-\gamma } \right]=i^{\gamma } E\left[X^{-\gamma } \right] 
\end{equation} 
with $Re\gamma >0$.

Summing up, in this first step we have shown that the classical relation between CF and moments (\ref{eq4}) is extended also to complex moments, by the RL definition of fractional operators. The next step is to use a representation of the CF (and of the PDF) in terms of its fractional derivatives of complex order calculated in zero. It will be demonstrated that fractional moments can be used in reconstructing the CF (or the PDF) of every distribution (also distributions with discontinuity in zero in the CF) extending the well known eq.(\ref{eq3}).

\section{Generalized Taylor series using fractional moments}
There are many generalizations of the Taylor series involving fractional derivatives, see e.g. [12], [15]--[21]. Riemann himself made the first attempt. In his posthumous published work he wrote
\begin{equation} 
\label{eq31} 
f\left(x+h\right)=\sum _{m=-\infty }^{\infty }\frac{h^{m+r} }{\Gamma \left(m+r+1\right)} \left({\rm {\mathcal D}}_{a_{+} }^{m+r} f\right)\left(x\right)  
\end{equation} 
with $r$ a fixed real number and $\left({\mathcal D}_{a_{+} }^{\gamma } f\right)\left(x\right)$ is the left--handed RL fractional derivative on a finite support 
\begin{equation} 
\label{eq32} 
\left({\rm {\mathcal D}}_{a+}^{\gamma } f\right)\left(x\right)=\frac{1}{\Gamma \left(n-\gamma \right)} \frac{d^{n} }{dx^{n} } \int _{a}^{x}\frac{f\left(\xi \right)d\xi }{\left(x-\xi \right)^{\gamma -n+1} }  ; x>a 
\end{equation} 
If $r$ in eq.(\ref{eq31}) is an integer, all the terms $m<-r$ disappear and it is equivalent to the classical Taylor series. However, Hardy [17] proved that the expansion (\ref{eq31}) is a divergent series. For $h$ large enough the series converges, but, as we are interested on the values of the derivatives of the CF in zero, the generalized Taylor expansion of Riemann is useless for our scope. Many other generalizations of the Taylor expansion using fractional derivatives exist. In Appendix B some of them are analyzed in detail showing that they have some undesired property for the representation of the CF. 

In this section an integral form of Taylor expansion related to the inverse Mellin transform, proposed by Samko et al.([12], pp.144--145)  will be applied to the CF and to the PDF of a random variable. Then, by means of eq.(\ref{eq21})--(\ref{eq25}), generalization of eq.(\ref{eq3}) will be provided. 

The starting point is to recognize that the RL integral (\ref{eq5}) may be interpreted as the Mellin transforms (see Appendix A) of the functions $f\left(x\pm \xi \right)$, with $x$ fixed, that is 
\begin{equation} 
\label{eq33} 
\Gamma \left(\gamma \right)\left(I_{\mp }^{\gamma } f\right)\left(x\right)=\int _{0}^{\infty }\xi ^{\gamma -1} f\left(x\pm \xi \right)d\xi   
\end{equation} 
Under this perspective, by performing the inverse Mellin transform, one gets 
\begin{equation} 
\label{eq34} 
f\left(x\pm \xi \right)=\frac{1}{2\, \pi \, i} \int _{\rho -i\infty }^{\rho +i\infty }\Gamma \left(\gamma \right)\left(I_{\mp }^{\gamma } f\right)\left(x\right)\left|\xi \right|^{-\gamma } d\gamma  ,   \rho =Re\gamma >0 
\end{equation} 
that for $x=0$ becomes 
\begin{equation}
\label{eq35} 
f\left(\pm \xi \right)=\frac{1}{2\, \pi \, i} \int _{\rho -i\infty }^{\rho +i\infty }\Gamma \left(\gamma \right)\left(I_{\mp }^{\gamma } f\right)\left(0\right)\left|\xi \right|^{-\gamma } d\gamma   
\end{equation} 

As claimed in [12], this integral may be interpreted as an integral form of the Taylor expansion in the sense that knowing $\left(I_{\pm }^{\gamma } f\right)\left(0\right)$ on some line $\rho =Re\left[\gamma \right]>0$, then the value of the function $f$ in the whole axis may be predicted. Under this consideration, eq.(\ref{eq35}) is the generalized Taylor expansion proposed by Samko et al.

\section{Integral representation of the CF in terms of complex moments}

Let us suppose that $\left(I_{\pm }^{\gamma } \phi _{X} \right)\left(0\right)$, with $\gamma \in {\bf {\mathbb C}}$ and $\rho =Re\gamma >0$, exists. In the following, conditions of existence will be given. Then, particularizing eq.(\ref{eq35}) and using eq.(\ref{eq25}) for the CF one obtains
\begin{eqnarray} 
\label{eq36}
\phi _{X} \left(\mp \vartheta \right)=\frac{1}{2\, \pi \, i} \int _{\rho -i_{} \infty }^{\rho +i_{} \infty }\Gamma \left(\gamma \right)\left(I_{\pm }^{\gamma } \phi _{X} \right)\left(0\right)\left|\vartheta \right|^{-\gamma } d\gamma  =\\ \nonumber
\, \, \, \, \, \, \, \, \, \, \, \, \, \, \, =\frac{1}{2\, \pi \, i} \int _{\rho -i_{} \infty }^{\rho +i_{} \infty }\Gamma \left(\gamma \right)E\left[\left(\pm iX\right)^{-\gamma } \right]\left|\vartheta \right|^{-\gamma } d\gamma  ;       
\end{eqnarray} 

The line integral in eq.(\ref{eq36}) can be rewritten as
\begin{equation} 
\label{eq37} 
\phi _{X} \left(\mp \vartheta \right)=\frac{1}{2\, \pi } \int _{-\infty }^{\infty }\Gamma \left(\rho +i_{} \eta \right)E\left[\left(\pm iX\right)^{-\rho -i_{} \eta } \right]\left|\vartheta \right|^{-\rho -i_{} \eta } d\eta   
\end{equation} 

Some remarks are necessary at this point: (i) due to the position $\rho >0$, the moments $E\left[\left(\mp iX\right)^{-\rho -i\eta } \right]$ in eq.(\ref{eq37}) have intrinsically negative real order; (ii) random variables with divergent integer moments do have, at least in an open interval, negative real order moments and can be represented by eq.(\ref{eq37}); (iii) eq.(\ref{eq36}) can be evaluated for every value of $\rho $ inside an interval, that is $0<\rho <1$, as we prove in the following.

For clarity's sake, we develop an easy example by considering the CF of a Gaussian distribution showing the relation between the generalized integral representation (\ref{eq36}) proposed and the Taylor series (\ref{eq3}). We recall that the characteristic function of a standard Gaussian random variable $X$ is $\phi _{X} \left(\vartheta \right)=e^{-\vartheta ^{2} /2} $. In order to apply eq.(\ref{eq36}) the value of $\rho =Re\gamma $ must be properly selected. As first consideration, from the definition of the RL fractional integral, the operator $\left(I_{\pm }^{\gamma } \phi _{X} \right)\left(\vartheta \right)$ is meaningful with $Re\gamma >0$. In addition, in case of Gaussian r.v., moments of the type $E\left[\left(\mp iX\right)^{-\gamma } \right]\, \, =\smallint _{-\infty }^{\infty } p_{X} \left(x\right)\left(\mp ix\right)^{-\gamma } dx$ exist in the range $-\infty <Re\gamma <1$. Combining the two conditions of existence, one obtains the so called \textit{fundamental strip }$0<Re\gamma <1$, that is an open interval in which eq.(\ref{eq36}) holds. It will be shown that every distribution can be represented by the integral representation (\ref{eq36} if $Re\gamma $ belongs to the range $0<Re\gamma <1$. Further, the integrand function is holomorph inside the fundamental strip, ensuring that one can choose \textit{every value} inside the fundamental strip. Outside the fundamental strip the integrand is not necessarily holomorph and shows isolated singularities in the no--positive part of the real axes. In  Figure 1, just the real part of the integrand is plotted choosing a particular value of $\vartheta =\tilde{\vartheta }>0$. 
%

It has to be noted that, in this particular case, the function has poles at 0,--2,--4,--6,...etc. Now, due to the residue theorem, integrating along the line with real abscissa $0<\rho <1$, corresponds to integrate in the whole half plane at the left of the line, that is to the sum of the residues at the poles of the integrand, obtaining:
\begin{eqnarray} 
\label{eq38}
\phi _{X} \left(\vartheta \right)=\frac{1}{2\, \pi \, i} \int _{\rho -i_{} \infty }^{\rho +i_{} \infty }\Gamma \left(\gamma \right)E\left[\left(-iX\right)^{-\gamma } \right]\vartheta ^{-\gamma } d\gamma  = \\ \nonumber
\sum _{k=0,-2,...}^{-\infty }{\rm Res}\left(\Gamma \left(k\right)E\left[\left(-iX\right)^{-k} \right]\vartheta ^{-k} \right) = \\ \nonumber
=\sum _{k=0,2,...}^{\infty }\frac{\left(i\vartheta \right)^{k} }{k!} E\left[X^{k} \right]  
\end{eqnarray} 
where ${\rm Res}(\cdot )$ is the residue, and the property ${\rm Res}(\Gamma \left(-k\right))=\left(-1\right)^{k} /k!,\, \, \, k=0,1,2,...$has been used. The former expression is the connection between the Taylor series representation of the characteristic function and the integral representation proposed. 

Now, it is well--known that using the Taylor series in terms of integer moments one needs a closure scheme in order to properly truncate the sum to a finite number of terms. In the integral representation by means of complex moments it is easier to find in which interval the integrand does not contribute anymore to the value of the integral. Indeed, as Figure 1 highlights, the integral (\ref{eq37} can be approximated by truncating it in a range $\left[-\bar{\eta },\bar{\eta }\right]$ and then numerically evaluated. In the next section, performing a basic rectangle integration scheme it is shown that with a finite number of fractional moments one can approximate either the CF both the PDF of every distribution.

\subsection{Series approximation of the CF in terms of fractional moments}

The integral we want to evaluate numerically by means of a simple integration scheme is
\begin{equation}
\label{eq39} 
\phi _{X} \left(\pm \vartheta \right)=\frac{1}{2\, \pi \, } \int _{-\infty }^{+\infty }\Gamma \left(\rho +i\eta \right)E\left[\left(\mp iX\right)^{-\rho -i\eta } \right]\left|\vartheta \right|^{-\rho -i\eta } d\eta   
\end{equation} 
that, can be approximated by truncating the extremes 
\begin{equation} 
\label{eq40} 
\phi _{X} \left(\pm \vartheta \right)\cong \frac{1}{2\, \pi \, } \int _{-\bar{\eta }}^{\bar{\eta }}\Gamma \left(\rho +i\eta \right)E\left[\left(\mp iX\right)^{-\rho -i\eta } \right]\left|\vartheta \right|^{-\rho -i\eta } d\eta   
\end{equation} 
Then, setting $\eta =k_{} \Delta $, with $k\in {\bf {\mathbb N}}$ and $\Delta \in {\bf {\mathbb R}}_{+} $, and letting $\gamma _{k} =\rho +ik\Delta $, with $\rho >0$, and being $\bar{\eta }=m\, \Delta $ (see Figure 2, eq.(\ref{eq40}) may be rewritten in discrete approximated form as follows
\begin{equation}
\label{eq41}
\phi _{X} \left(\vartheta \right)\cong \frac{\Delta }{2\, \pi \, } \sum _{k=-m}^{m}\Gamma \, \left(\gamma _{k} \right)E\left[(-iX)^{-\gamma _{k} } \right]\left|\vartheta \right|^{-\gamma _{k} } \,\,\,\,\,\, {\rm for} \vartheta >0
\end{equation}

\begin{equation}
\label{eq42}
\phi _{X} \left(\vartheta \right)\cong \frac{\Delta }{2\, \pi \, } \sum _{k=-m}^{m}\Gamma \, \left(\gamma _{k} \right)E\left[(iX)^{-\gamma _{k} } \right]\left|\vartheta \right|^{-\gamma _{k} } \,\,\,\,\,\,  {\rm for} \vartheta <0
\end{equation}


Of course, CF has the property that $\phi _{X} \left(-\vartheta \right)$ is the complex conjugate of $\phi _{X} \left(\vartheta \right)$, in the following denoted as $\phi _{X}^{*} \left(\vartheta \right)$, and this fact will lead to the interesting simplification that only one equation between eqs.(\ref{eq41})--(\ref{eq42}) suffices to represent the whole CF.

By introducing eqs.
(\ref{eq29})--(\ref{eq30}) in (\ref{eq41})--(\ref{eq42}), the series representation of the CF in terms of moments and absolute moments may be rewritten in the form
\begin{equation} 
\label{eq43}
\phi _{X} \left(\vartheta \right)\cong \frac{\Delta }{2\, \pi \, } \sum _{k=-m}^{m}\Gamma \, \left(\gamma _{k} \right)i^{\gamma _{k} } E\left[X^{-\gamma _{k} } \right]\left|\vartheta \right|^{-\gamma _{k} }      
\end{equation} for $\vartheta >0$ 
\begin{eqnarray}
\label{eq44}
{\phi _X}\left( \vartheta  \right) = \frac{\Delta }{{2\pi }}\sum\limits_{k =  - \infty }^\infty  {\Gamma \left( {{\gamma _k}} \right)\left\{ {2\cos \left( {\frac{{\pi {\gamma _k}}}{2}} \right)E\left[ {{{\left| X \right|}^{ - {\gamma _k}}}} \right]} \right. + } \\ \nonumber
\,\,\,\,\,\,\,\,\,\,\,\,\,\,\,\,\,\,\,\,\,\,\,\,\,\,\,\,\,\,\,\,\,\,\,\,\,\,\,\,\,\,\,\,\,\,\,\,\,\,\,\,\,\,\,\,\,\,\,\,\,\,\,\,\,\,\,\,\,\,\,\,\,\left. { - {i^{{\gamma _k}}}E\left[ {{X^{ - {\gamma _k}}}} \right]{{\left| \vartheta  \right|}^{ - {\gamma _k}}}} \right\}
\end{eqnarray}
for $\vartheta <0$ 

Note that, since $\phi _{X} \left(\vartheta \right)=\phi _{X}^{*} \left(-\vartheta \right)$, eq.(\ref{eq43}) is enough in order to define the CF in all the range $\left]-\infty ,\infty \right[$. Furthermore, eq.(\ref{eq43}) has a mathematical form very similar to that of the classical Taylor expansion of the CF given in eq.(\ref{eq3}).

We investigate now for the bounds of $\rho $ such that this series representation is valid. Of course, due to the presence of the RL fractional integral, $\rho $ must be positive from definition (\ref{eq5}), so, the first lower bound of $\rho $ is zero. In order to have a higher bound, one shall recall the condition of existence of the inverse Mellin transform. The choice of the value of $\rho $, indeed, must belong to the so called fundamental strip of the Mellin transform  in eq.(\ref{eq36}), (see Appendix A). In order to define the fundamental strip of the Mellin transform, it suffices to note that the integral  

\begin{equation} 
\label{eq45} 
{\mathcal M}\left\{\phi _{X} \left(\pm \xi \right);\vartheta \right\}=\int _{0}^{\infty }\xi ^{\gamma -1} \phi _{X} \left(\pm \xi \right)d\xi   
\end{equation} 

converges, if the integral 
\begin{equation} 
\label{eq46} 
\left|\int _{0}^{\infty }\xi ^{\gamma -1} \phi _{X} \left(\pm \xi \right)d\xi  \right|<\int _{0}^{\infty }\xi ^{\rho -1} \left|\phi _{X} \left(\pm \xi \right)\right|d\xi   
\end{equation} 
at the r.h.s. converges. As the characteristic function is absolutely convergent in the real axis, eq.(\ref{eq46}) is bounded at least for every $0<\rho <1$. Then, in order to use the integrals (\ref{eq39})--(\ref{eq40}) or the series (\ref{eq41})--(\ref{eq42}), it suffices to choose a value of $\rho $ in the interval $0<\rho <1$. This represents the strictest condition of existence of the series proposed. Of course, there are many distributions satisfying (\ref{eq45}) in a wider fundamental strip. For example, the CF of a standard Gaussian distribution admits a Mellin transform in the fundamental strip $0<\rho <\infty $ and we might choose every $\rho >0$ in evaluating eq.(\ref{eq39}).

Suppose we want to use eq.(\ref{eq43}) for the representation of the CF of an $\alpha$--stable random variable. It is known that for such distributions moments of the type $E\left[X^{-\rho -i\eta } \right]$ are finite, although complexes, only if $-1<-\rho <\alpha $, otherwise they diverge due to the heavy tails of the PDF. Being the stability index $\alpha $ defined in the range $0<\alpha \le 2$, the condition $0<\rho <1$ on the fundamental strip allows to represent the CF by means of such moments $E\left[X^{-\rho -i\eta } \right]$ with intrinsically negative real order $-1<-\rho <0$. 

In order to understand the role played by the parameter $\Delta $, we first recall that the series form representation is the numerical integration of eq.(\ref{eq36}). For fixed $\rho $, the integration is performed in the imaginary axis as shown in Figure 2. Small values of $\Delta $ produces high accuracy, however since from a numerical point of view in any cases the summation (\ref{eq43}) has to be truncated retaining a certain number, say $m$ number of terms, if $\Delta $ is very small then $m_{} \Delta =\bar{\eta }$ remains small and we may exclude in the integration significant part of the integrand. Then, $\bar{\eta }$ must be chosen in a such way that $\Gamma \left(\rho +i\bar{\eta }\right)E\left[\left(\mp iX\right)^{-\rho -i\bar{\eta }} \right]\left|\vartheta \right|^{-\rho -i\bar{\eta }} $  is negligible.

\subsection{Probability density function representation in terms of fractional moments}

In engineering problems, i.e. in solution of stochastic differential equations or in path integral methods, we are concerned more in probability, rather than in its spectral counterpart. In this section it is derived a dual integral representation of the PDF in terms of complex moments, by means of Fourier transform. 

The inverse Fourier transform of the CF can be performed by ${\rm {\mathcal F}}^{-1} \left\{\phi _{X} \left(\vartheta \right);x\right\}$ or, that is the same, ${\rm {\mathcal F}}^{-1} \left\{\phi _{X} \left(\vartheta >0\right)+\phi _{X} \left(\vartheta <0\right);x\right\}$ by the linearity of the integration involved. Then, the inverse Fourier transform of relation (\ref{eq36}) may be performed in the following way
\begin{eqnarray} 
\label{eq47}
\\ \nonumber
p_{X} \left(x\right)={\rm {\mathcal F}}^{-1} \left\{\phi _{X} \left(\vartheta \right);x\right\}=\frac{1}{\left(2\pi \right)^{2} \, i} \int _{\rho -i_{} \infty }^{\rho +i_{} \infty }\Gamma \left(\gamma \right)\left\{\left(I_{+}^{\gamma } \phi _{X} \right)\left(0\right)\int _{0}^{\infty }\vartheta ^{-\gamma } \exp \left(-i\vartheta x\right)d\vartheta  +\right. \\ \nonumber
\, \, \, \, \, \, \, \, \, \, \, \, \, \, \, \, \, \, \, \, \, \, \, \, \, \, \, \, \, \, \, \, \, \, \, \, \, \, \, \, \, \, +\left(I_{-}^{\gamma } \phi _{X} \right)\left(0\right)\left. \int _{-\infty }^{0}\left(-\vartheta \right)^{-\gamma } \exp \left(-i\vartheta x\right)d\vartheta  \right\}d\gamma  
\end{eqnarray} 
that, under the condition $0<\rho <1$ can be further simplified

\begin{equation} 
\label{eq48} 
\\ \nonumber
p_{X} \left(x\right)=\frac{1}{\left(2\pi \right)^{2} \, i} \int _{\rho -i_{} \infty }^{\rho +i_{} \infty }\Gamma \left(\gamma \right)\Gamma \left(1-\gamma \right)\left\{\left(I_{+}^{\gamma } \phi _{X} \right)\left(0\right)\left(ix\right)^{\gamma -1} +\left(I_{-}^{\gamma } \phi _{X} \right)\left(0\right)\left(-ix\right)^{\gamma -1} \right\} \, d\gamma  
\end{equation} 
Finally, substituting eq.(\ref{eq25}) in the latter equation, the relation searched reads
\begin{equation} 
\label{eq49} 
\\ \nonumber
p_{X} \left(x\right)=\frac{1}{\left(2\pi \right)^{2} \, i} \int _{\rho -i_{} \infty }^{\rho +i_{} \infty }\Gamma \left(\gamma \right)\Gamma \left(1-\gamma \right)\left\{E\left[\left(-iX\right)^{-\gamma } \right]\left(ix\right)^{\gamma -1} +E\left[\left(iX\right)^{-\gamma } \right]\left(-ix\right)^{\gamma -1} \right\} \, d\gamma  
\end{equation} 
and it is the integral form of the Taylor approximation of the PDF in terms of moments.

It has to be remarked that, since $\phi _{X} \left(\vartheta \right)=\phi _{X}^{*} \left(-\vartheta \right)$, then $p_{X} \left(x\right)$ may be also rewritten in a more compact form as follows
\begin{eqnarray} 
\label{eq50}
p_{X} \left(x\right)=\frac{1}{2_{} \pi ^{2} i} Re\left\{\int _{\rho -i_{} \infty }^{\rho +i_{} \infty }\Gamma \left(\gamma \right)\Gamma \left(1-\gamma \right)\left(I_{+}^{\gamma } \phi _{X} \right)\left(0\right)\left(ix\right)^{\gamma -1}  \, d\gamma \right\}= \\ \nonumber
\, =\frac{1}{2_{} \pi ^{2} i} Re\left\{\int _{\rho -i_{} \infty }^{\rho +i_{} \infty }\Gamma \left(\gamma \right)\Gamma \left(1-\gamma \right)E\left[\left(-iX\right)^{-\gamma } \right]\left(ix\right)^{\gamma -1}  \, d\gamma \right\}\,  
\end{eqnarray} 
Then, with the same set of moments $E\left[\left(-iX\right)^{-\gamma } \right]$ one can represent both the CF, with eq.(\ref{eq36}) evaluated only for $\vartheta >0$, or the PDF evaluated by eq.(\ref{eq50}), without additional calculation (i.e. Fourier transform).

Expression (\ref{eq50}) can be approximated in the form
\begin{equation}
\label{eq51} 
p_{X} \left(x\right)\cong \frac{\Delta }{2\pi ^{2} } Re\left\{\sum _{k=-m}^{m}\Gamma \left(\gamma _{k} \right)\Gamma \left(1-\gamma _{k} \right)E\left[\left(-iX\right)^{-\gamma _{k} } \right]\left(ix\right)^{\gamma _{k} -1}  \right\} 
\end{equation} 
Taking into account eq.(\ref{eq30}), previous expression simplifies 

\begin{equation} 
\label{eq52} 
p_{X} \left(x\right)\cong \frac{\Delta }{2_{} \pi ^{2} } Re\left\{\sum _{k=-m}^{m}\Gamma \left(\gamma _{k} \right)\Gamma \left(1-\gamma _{k} \right)i^{\gamma _{k} } E\left[X^{-\gamma _{k} } \right]\left(ix\right)^{\gamma _{k} -1}  \right\} 
\end{equation} 
Such a perfect duality between the two representations in $\vartheta $ and $x$ domains, evident comparing eqs. (\ref{eq43}) and (\ref{eq52}), may not be evicted working in terms of classical moments. In fact an inverse Fourier transform of (\ref{eq3}) gives the PDF $p_{X} \left(x\right)$ in the following form

\begin{equation}
\label{eq53} 
p_{X} \left(x\right)=\sum _{j=0}^{\infty }\left(-1\right)^{j} \frac{E\left[X^{j} \right]}{j!} \frac{d^{j} \delta \left(x\right)}{dx^{j} }   
\end{equation} 
that cannot be if practical use in representing the PDF because of the Dirac's delta $\delta \left(x\right)$ derivatives. By summing up knowledge of moments of the type $E\left[X^{-\gamma _{k} } \right]$, $\gamma _{k} \in {\bf {\mathbb C}}$ are able to represent both the CF and the PDF in the whole $\vartheta $ and $x$ domains.

\section{Numerical results}
In this section, we present some simple applications in order to show the validity of eq.(\ref{eq21}) for both the cases in which the CF is differentiable in zero and for case of $\alpha$--stable random variables. In the following, $\rho $ will be used to indicate the real part of $\gamma $. As it has been already pointed out, the value of $\rho $ inside the fundamental strip can be arbitrarily chosen, then the results here reported for particular values of $\rho $ can be reproduced of course with different choice of its value.
Let $X$ be a random variable with uniform distribution in $\left[-a,a\right]$, then all the moments exist and are given as 
\begin{equation} 
\label{eq54} 
E\left[X^{2k} \right]=\frac{a^{2k} }{2k+1} ;  E\left[X^{2k+1} \right]=0, k=1,2,... 
\end{equation} 

while the CF is 
\begin{equation} 
\label{eq55} 
\phi _{X} \left(\vartheta \right)=\frac{\sin \left(\vartheta ^{} a\right)}{\vartheta _{} a}  
\end{equation} 
The classical Taylor expansion of the CF is then 
\begin{equation} 
\label{eq56} 
\phi _{X} \left(\vartheta \right)=\sum _{k=0}^{\infty }\frac{a^{2k} }{\left(2k+1\right)\left(2_{} k\right)!}  \, \left(i\vartheta \right)^{2k}  
\end{equation} 
Such a series converges uniformly to $\phi _{X} \left(\vartheta \right)$ in the entire domain. However, when a truncation is performed the CF diverges as it is shown in Figure 3a. The latter circumstance produces an undesirable result when the PDF of $X$ has to be restored by Fourier transform of the truncated series (\ref{eq56}). The converse happens by using the fractional series proposed. In fact, as shown in Figure 3b, by applying (\ref{eq41})--(\ref{eq42}) to the CF (\ref{eq55}), one notes that the fractional series remains always convergent in the whole domain and for $\vartheta \to \infty $, $\phi _{X} \left(\vartheta \right)\to 0$. 
Easy calculations give the moments of complex order $\left(-\gamma \right)$ for the uniform distribution, i.e. $E\left[\left(\pm iX\right)^{-\gamma } \right]=\left(ia\right)^{-\gamma } /(1-\gamma )$ defined in a strip  $-\infty <\rho <1$; in particular, in Figure 3b it has been assumed that $\rho =0.4$ and $\Delta =0.4$. 


In order to show that the proposed series works also for complex CF, in Figure 4, the proposed expansion of the CF is reported for a Rayleigh distribution with scale [da vedere]. In Figure 4a and 4b the real part and the imaginary part are plotted, respectively. The complex moments of a Rayleigh distribution with scale parameter $\sigma \in {\bf {\mathbb R}}>0$ are given by $E\left[\left(\mp iX\right)^{-\gamma } \right]=2^{-\gamma /2} \left(\mp i\sigma \right)^{-\gamma } \Gamma \left(1-\gamma /2\right)$, within the strip $-\infty <\rho <2$. Then, in this case one could choose the fundamental strip in the interval $0<\rho <2$. Despite, the knowledge of fractional moments in some line inside the strip $0<\rho <1$ ensures that the series (\ref{eq41})--(\ref{eq42}) is meaningful. The results of the fractional series  reported in Figure 4 have been calculated with the choice of $\rho =0.4$, but similar results may be obtained also for other values in the range of validity of $0<\rho <1$, and $\Delta =0.4$.


A more interesting and striking example is the symmetric Cauchy random variable $X$ with $p_{X} \left(x\right)=\pi ^{-1} /(x^{2} +1)$. For such a distribution the moments exist only for $-1<\rho <1$, then the usual Taylor expansion fails because it would be just a sum of divergent terms, and therefore meaningless. Further, in this case it may be shown that moments $E\left[\left(\mp iX\right)^{-\gamma } \right]=1$, in the strip $-1<\rho <1$,  and the approach in this paper proposed using eqs. (\ref{eq41}) and (\ref{eq42}), is a very good approximation, as shown in Figure 5.


%
%
%

Then, we remark that the fractional series representation works also for distributions with moments existing only in a limited range, like the stable distributions, because one needs only the existence of a real moment $E\left[X^{-\rho } \right]$ in, at least, one real value $\rho $, such that $0<\rho <1$. 

To enforce this concept, as last example, the L\'evy stable variable with PDF given by \\ $p_{X} \left(x\right)=\left(2\pi \right)^{-1/2} \left(x\right)^{-3/2} \exp \left(-1/2x\right)$, whose moments $E\left[\left(-iX\right)^{-\gamma } \right]$ exist if the condition $-1/2<\rho <\infty $ is satisfied. The CF of this distribution has been approximated with $\rho =0.9$, then moments of the type $E\left[X^{-0.9-i\eta } \right]$, have been used in series (\ref{eq43})--(\ref{eq44}) or, that is the same, moments $E\left[\left(\mp iX\right)^{-0.9-i\eta } \right]$ in (\ref{eq41})--(\ref{eq42}). Figure 6a and Figure 6b show the convergence of the series proposed.

Now, we report some interesting applications of the fractional series eq.(\ref{eq51}) to find the PDF from the knowledge of some moments, that is, without using inverse Fourier transform. In Figure 7 one can see that the PDF of a Gaussian random variable is well approximated by 30 complex moments in the whole domain, having used $\rho =0.4$ and $\Delta =0.4$. Figure 8 and Figure 9  show that in order to represent the PDF of a Cauchy and a L\'evy random variables respectively, 10 complex moments are sufficient, with the choice of $\rho =0.4$ and $\Delta =0.4$.

\section{Conclusions}
In this paper, a new perspective of the probabilistic characterization of the random variables has been highlighted. It has been shown that the fractional derivatives of the characteristic function in zero coincide with the fractional moments. Then, integral Taylor approximation due to Samko et al. [12] has been the starting point to develop a representation of both the density and the characteristic function of every random variable. Such a new representation offers many advantages: (i) it may be applied to not--differentiable in zero characteristic functions, and to densities with heavy--tails; (ii) with a unique set of moments $E\left[\left(-iX\right)^{-\gamma } \right]$ both the PDF and the CF can be represented in dual form; (iii) the series proposed are not of integer power--type and once truncated they vanish at infinity.

The authors have recently extended to multivariate random variables the representation proposed in this paper. Moreover, a path integral method for the solution of non--linear one--dimensional stochastic differential equation excited by white noise processes based on complex moments has been studied (see [22] and [23]), while the solution of multidimensional stochastic differential equation is underway by the authors. Indeed, the path integral method, which is based on the evolution in time of the response density, has proven to take full advantage from the PDF representation in terms of complex moments presented in this paper. 

\section{Appendix A}

We gather in this appendix some well--known definitions of the Mellin transform needed in the text. Further, we briefly describe the so called composition rules of fractional operators.

The Mellin transform ([12], p.25; [24], p.41) of a function $f\left(x\right)$,$x>0$, $s\in {\bf {\mathbb C}}$, is defined as $\varphi \left(s\right){\rm }=\, {\rm {\mathcal M}}\left\{f\left(x\right);s\right\}=\smallint _{0}^{\infty } f\left(\xi \right)\, \xi ^{\, s{\rm -1}} d\xi $ along the inverse operator ${\rm }f\left(x\right){\rm }=\, {\rm {\mathcal M}}^{\, -{\rm 1}} \left\{\varphi \left(s\right);x\right\}$ $=\left(2\pi i\right)^{-1} \smallint _{c-i\infty }^{c+i\infty } \varphi \left(s\right)\, x^{-s} ds$, with $\, c=Re\left(s\right)$.  The Mellin transform $\varphi \left(s\right)$ of the function $f\left(x\right)$, such that 
\begin{equation}
\label{eqA1} 
f\left(x\right)=\left\{\begin{array}{c} {O\left(x^{p} \right),\, \, x\to 0} \\ {O\left(x^{q} \right),\, \, x\to \infty } \end{array}\right.  
\end{equation} 
exists, and is analytic, in the \textit{fundamental strip} $-p<Re\left[s\right]<-q$. In order to perform the inverse Mellin transform $c=Res$ must be chosen inside the fundamental strip. Tables of Mellin transforms of commonly used functions are given in [24] and [25]. 

Properties of nested application of the fractional operators, called composition rules, are useful in calculations. Suppose the existences of the integrals and derivatives involved in the following relations, and consider $Re\gamma _{1} >0$ and $Re\gamma _{2} >0$. Then, the properties 
\begin{equation} 
\label{eqA2} 
I_{+}^{\gamma _{1} } I_{+}^{\gamma _{2} } f=I_{+}^{\gamma _{1} +\gamma _{2} } f=I_{+}^{\gamma _{2} } I_{+}^{\gamma _{1} } f 
\end{equation} 
\begin{equation} 
\label{eqA3} 
I_{-}^{\gamma _{1} } I_{-}^{\gamma _{2} } f=I_{-}^{\gamma _{1} +\gamma _{2} } f=I_{-}^{\gamma _{2} } I_{-}^{\gamma _{1} } f 
\end{equation} 

on the composition of different order integrals, are valid. On the converse, composition rule between fractional derivative and fractional integral does not follow straightforwardly. Indeed, to better understand the commutation property between fractional integration and differentiation some parallel with the standard differential calculus might help. In ordinary calculus, it is well--known that performing firstly the integral of a function and then a derivative produces the original function, i.e. $\frac{d}{dx} \int _{a}^{x}f\left(x\right)dx=f\left(x\right) $. The same happens to \textit{n}--order derivatives and \textit{n}--folded integrals. Exactly in the same way, applying the fractional operators in the order ${\rm {\mathcal D}}_{}^{\rho } I_{}^{\rho } f$ gives the original function $f$. 

Instead, performing firstly the derivative of $f$ and then integrating, i.e. $\int _{a}^{x}\frac{d}{dx} f\left(x\right)dx $, one obtains the function plus a constant, or for \textit{n}--order derivatives and \textit{n}--folded integrals  one obtains the function $f$ plus a polynomial of order $n-1$. This simple argument is valid also in case of fractional operators and consequently the relation $I_{}^{\rho } {\rm {\mathcal D}}_{}^{\rho } f=f$ is not in general true, but depends on the function $f\left(x\right)$. It has been shown ([12], p. 43--45) that the latter is true only for those functions vanishing at the interval boundaries.

\section{Appendix B}

Although many generalization of the Taylor series exist (see e.g. [15] -- [21]), the only series adapt to express the CF in terms of moments is the series proposed, descending from the Mellin transform. In order to avoid useless attempts to readers interested on this topic, we describe in this short appendix some reasoning on other possible ways, at least at first sight, to develop a CF representation by means of a generalized Taylor expansion.

The first family of generalized Taylor series we examine descend from the composition rule of fractional operators aforementioned. In order to simplify the discussion we will use the left--handed RL fractional integral and derivative on a finite support defined as

\begin{equation}
\label{eqB1} 
\left(I_{a+}^{\gamma } f\right)\left(x\right)=\frac{1}{\Gamma \left(\gamma \right)} \int _{a}^{x}\frac{f\left(\xi \right)d\xi }{\left(x-\xi \right)^{1-\gamma } }  ;            x>a 
\end{equation} 

\begin{equation} 
\label{eqB2} 
\left({\rm {\mathcal D}}_{a+}^{\gamma } f\right)\left(x\right)=\frac{1}{\Gamma \left(n-\gamma \right)} \frac{d^{n} }{dx^{n} } \int _{a}^{x}\frac{f\left(\xi \right)d\xi }{\left(x-\xi \right)^{\gamma -n+1} }  ;   x>a 
\end{equation}

It has been already stressed that, in general, the fractional derivatives and integrals do not commute, if the derivative is performed as first operation, that is 
\begin{equation} 
\label{eqB3} 
I_{a+}^{\gamma } {\rm {\mathcal D}}_{a+}^{\gamma } f\ne f\left(x\right),       Re\gamma >0 
\end{equation} 
but, it has been proved in [12], that (\ref{eqB3}) is replaced by 
\begin{equation}
\label{eqB4} 
I_{a_{+} }^{\gamma } {\mathcal D}_{a_{+} }^{\gamma } f=f\left(x\right)-\sum _{k=0}^{n-1}\frac{\left(x-a\right)^{\gamma -k-1} }{\Gamma \left(\gamma -k\right)} D^{n-k-1}  I_{a_{+} }^{1-\left\{\gamma \right\}} \left(a\right),        
\end{equation} 
where the operator $D^{j} $ is the classical derivative of order \textit{jth}. Rearranging this relation one find an analogue of the Taylor theorem, expressing the function $f\left(x\right)$ as sum of fractional derivatives calculated in a point 

\begin{equation} 
\label{eqB5} 
f\left(x\right)=\sum _{k=-n}^{n-1}\frac{\left({\mathcal D}_{a_{+} }^{\gamma +k} f\right)\left(a\right)}{\Gamma \left(\gamma +k+1\right)}  \left(x-a\right)^{\gamma +k} +I_{a_{+} }^{\gamma +n} {\mathcal D}_{a_{+} }^{\gamma +n} f,        
\end{equation} 
with $n=\left[Re\gamma \right]+1$. From this formula one can state that, if for a function the property $I_{a_{+} }^{\gamma } {\mathcal D}_{a_{+} }^{\gamma } f=f\left(x\right)$ holds, then the function cannot be represented by the fractional Taylor series (\ref{eqB5}), because all the terms of the sum actually vanish. In the perspective of this paper, in order to understand if relation (\ref{eqB5}) is useful, it suffices to study the operation $I_{a_{+} }^{\gamma } {\mathcal D}_{a_{+} }^{\gamma } \phi _{X} \left(\vartheta \right)$. Recalling that the CF of a random variable $X$ is $\phi _{X} \left(\vartheta \right)=E\left[\exp \left(i\vartheta x\right)\right]$, or, that is the same, $I_{a_{+} }^{\gamma } {\mathcal D}_{a_{+} }^{\gamma } \phi _{X} \left(\vartheta \right)=I_{a_{+} }^{\gamma } {\mathcal D}_{a_{+} }^{\gamma } E\left[\exp \left(i\vartheta x\right)\right]$=$E\left[I_{a_{+} }^{\gamma } {\mathcal D}_{a_{+} }^{\gamma } \exp \left(i\vartheta x\right)\right]$. By means of direct calculations it is easy to verify that $I_{0_{+} }^{\gamma } {\mathcal D}_{0_{+} }^{\gamma } \exp \left(i\vartheta x\right)=\exp \left(i\vartheta x\right)$ and therefore series (\ref{eqB5}) is useless to our scope, with $a=0$. Other values of $a$ would not produce a connection between CF and moments. Same reasoning applies to the generalized Taylor series proposed by Dzherbashyan and Nersesyan [19].

Other Taylor expansions have been given by Jumarie [20]
\begin{equation} 
\label{eqB6} 
f\left(x\right)=\sum _{k=0}^{\infty }\frac{x^{\rho _{} k} }{\Gamma \left(\rho _{} k+1\right)}  \left({\rm {\mathcal D}}_{0+}^{\rho _{} k} f\right)\left(0\right),  0<\rho <1 
\end{equation} 

or, for $z\in {\bf {\mathbb C}}$, by Osler [16]
\begin{equation} 
\label{eqB7} 
f\left(z\right)=\rho \sum _{k=-\infty }^{\infty }\frac{\left(z-z_{0} \right)^{\rho _{} k} }{\Gamma \left(\rho _{} k+1\right)}  \left({\rm {\mathcal D}}_{0+}^{\rho _{} k} f\right)\left(z_{0} \right),        
\end{equation} 
Using relations (\ref{eqB6}), (\ref{eqB7}) it was not possible to extend the eq. (\ref{eq3}) for two reasons. Firstly, it was not possible to find an explicit relation between the fractional derivatives $\left({\rm {\mathcal D}}_{0+}^{\rho _{} k} \phi _{X} \right)\left(\vartheta \right)$ of the CF and the fractional moments, being $\left({\rm {\mathcal D}}_{0+}^{\rho _{} k} \phi _{X} \right)\left(\vartheta \right)$ defined just for values of $\vartheta \ge 0$, since the Fourier transform of such operator, like (\ref{eq12}), does not exist to our knowledge. Moreover, even if such an expression would exist, terms of the kind $\left({\rm {\mathcal D}}_{0+}^{\rho _{} k} \phi _{X} \right)\left(\vartheta \right)$, would involve fractional moments whose real part increases with $k=...-1,0,1,...$ and therefore not applicable to CF not differentiable in zero.

\section*{References}

\begin{enumerate}
\item \textbf{ }Samorodnitsky G, Taqqu MS. Stable non--Gaussian random processes: Stochastic models with infinite variance. Chapman and Hall, New York, 1994.

\item  Grigoriu M. Applied non--Gaussian processes: examples, theory, simulation, linear random vibration, and Matlab solutions. Prentice Hall, Englewoods Cliffs, StateNJ, 1995.

\item  Hilfer R. (Ed.). Applications of fractional calculus in physics, World Scientific Publishing Co, 2000.

\item  Carpinteri A, Chiaia B, Cornetti P. On the mechanics of quasi--brittle materials with a fractal microstructure. Engineering Fracture Mechanics 2003; 70: 2321--2349.

\item  Di Paola M, Zingales M. Long--range cohesive interactions of non--local continuum faced by fractional calculus. International Journal of Solids and Structures 2008; 45(21): 5642--5659

\item  Cottone G, Di Paola M, Zingales M. Fractional mechanical model for the dynamics of non--local continuum. Advances in Numerical Methods. Volume of the series: Lecture Notes in Electrical Engineering, (Eds. Mastorakis N, Sakellaris J). Springer--Verlag.

\item  Chechkin A, Gonchar V, Klafter J, Metzler R, Tanatarov L. Stationary states of non--linear oscillators driven by L\'evy noise. Chemical Physics 2002; 284: 233--251.

\item  Spanos PD, Zeldin BA. Random vibration of systems with frequency--dependent parameters or fractional derivatives. ASCE Journal of Engineering Mechanics, 1997; 123: 290--2.

\item  Metzler R, Klafter J. The random walk's guide to anomalous diffusion: a fractional dynamics approach. Physics Reports 2000; 339: 1--77.

\item  Metzler R, Klafter J. The restaurant at the end of the random walk: recent developments in the description of anomalous transport by fractional dynamics. Journal of Physics A: Mathematical and General 2004; 37: R161--R208.

\item  Wolfe S. On moments of probability distribution function, in Ross B, Fractional Calculus and Its Applications. placeCitySpringer--Verlag, StateBerlin, 1975.

\item  Samko GS, Kilbas AA, Marichev OI. Fractional integrals and derivatives, Theory and Applications, Gordon and Breach Science Publishers, 1993.

\item  Miller KS, Ross B. An introduction to the fractional calculus and fractional differential equations. John Wiley \& Sons, StateplaceNew York, 1993.

\item Oldham KB and Spanier J. The fractional calculus. Academic Press, StateplaceNew York, 1974.

\item  Riemann B. Versuch einer allgemeinen Auffassung der Integration und Differentiation. Gesammelte Math. Werke und Wissenschaftlicher. Teubner, Leipzig, 1876: 331--334.

\item  Osler T. An Integral Analogue of Taylor's Series and Its Use in Computing Fourier Transforms, Mathematics of Computation. 1972; 26: 449--460.

\item  Hardy G. Riemann's form of Taylor's series, J. London Math. Soc. 1945; 20: 48--57.

\item  Bertrand J, Bertrand P, Ovarlez JP. The Mellin Transform. The transforms and Application Handbook. Ed. A.D. Poularikas, Volume of the Electrical Engineering Handbook series, CRC press, 1995: Chapter 12.

\item  Dzherbashyan M, Nersesyan A. The criterion of the expansion of the functions to the Dirichlet series (in Russian), Izv. Akad. Nauk Armyan. SSR, Ser. Fiz.--Mat. Nauk. 1958; 11: 5, 85--108.

\item  Jumarie G. Modified Riemann--Liouville derivative and fractional Taylor series of nondifferentiable functions further results. Computer and Mathematics with Applications. 2006; 51: 1367--1376.

\item Trujillo J, Rivero M, Bonilla B. On a Riemann--Liouville Generalized Taylor's Formula, Journal of Mathematical Analysis and Applications. 1999; 231: 255--265.

\item  Cottone G, Di Paola M, Pirrotta A. Fractional moments and path integral solution for non linear systems driven by normal white noise. Proceedings of the International Symposium on Recent Advances in Mechanics, Dynamical Systems and Probability Theory,MDP -- 2007 Palermo, June 3--6, 2007. 

\item  Cottone G, Di Paola M, Pirrotta A. Path integral solution by fractional calculus. Journal of Physics: Conference Series. 2008; vol. 96: 1--11. doi:10.1088/1742--6596/96/1/012007.

\item  Sneddon IN. Fourier Transform. McGraw--Hill Book Company, 1951.

\item  Marichev OI. Handbook of integral transform of higher functions: Theory and algorithmic tables. Ellis Horwood Limited, 1982. 
\end{enumerate}

\begin{figure}[p]
\centering
\includegraphics[width=8cm,angle=270]{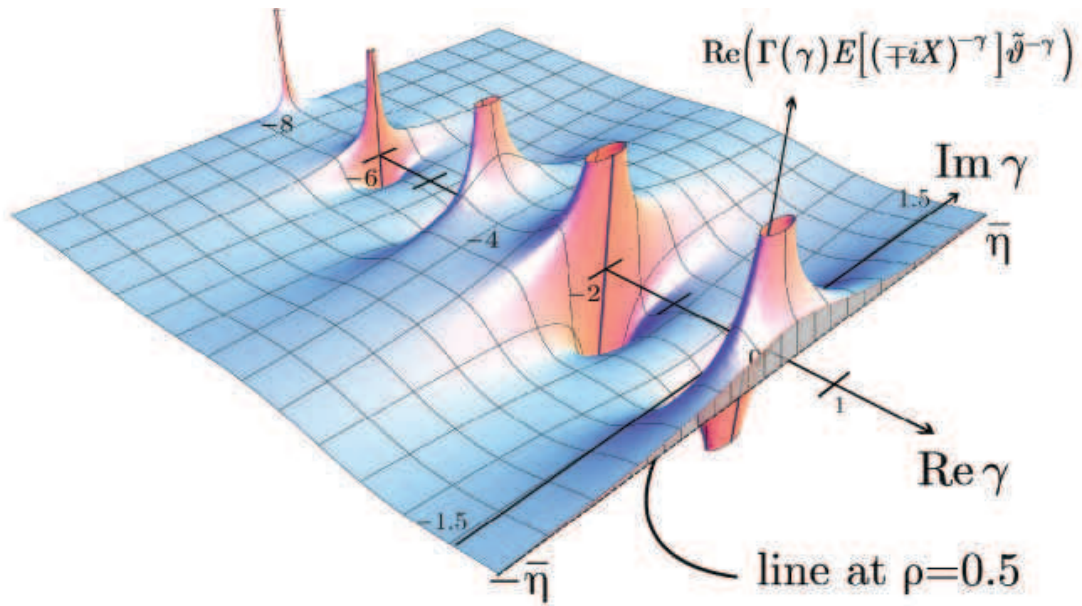}
\caption{Plot of the real part of the integrand in eq.(\ref{eq36}) for $\vartheta =\tilde{\vartheta }>0$.}
\label{fig1}
\end{figure}

\begin{figure}[p]
\centering
\includegraphics[width=8cm]{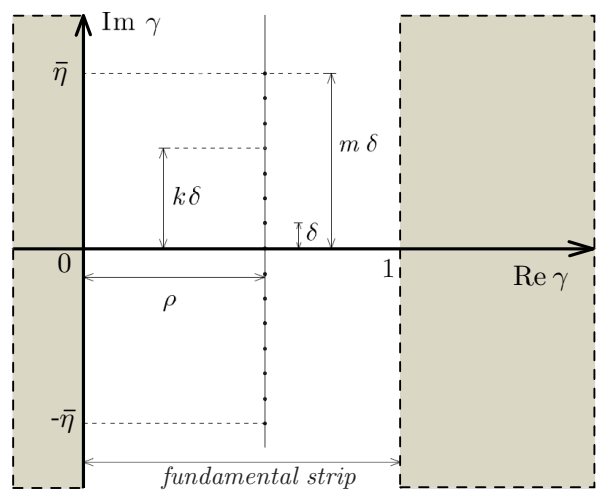}%
\caption{Numerical integration scheme in the fundamental strip.}%
\label{fig2}%
\end{figure}

\begin{figure}[p]
 \centering
 \subfigure[]
   {\includegraphics[width=6.5cm]{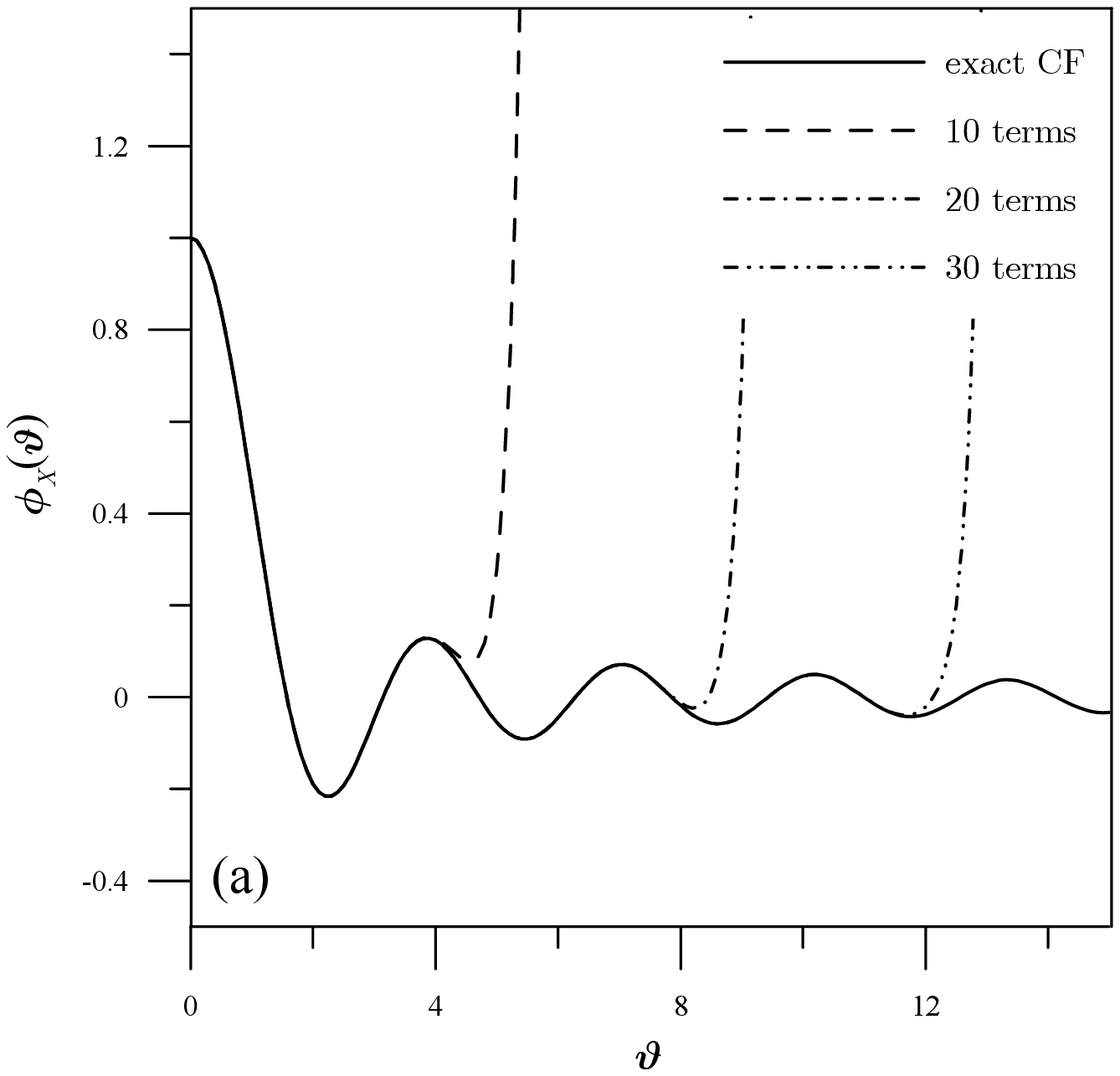}}
 \subfigure[]
   {\includegraphics[width=6.5cm]{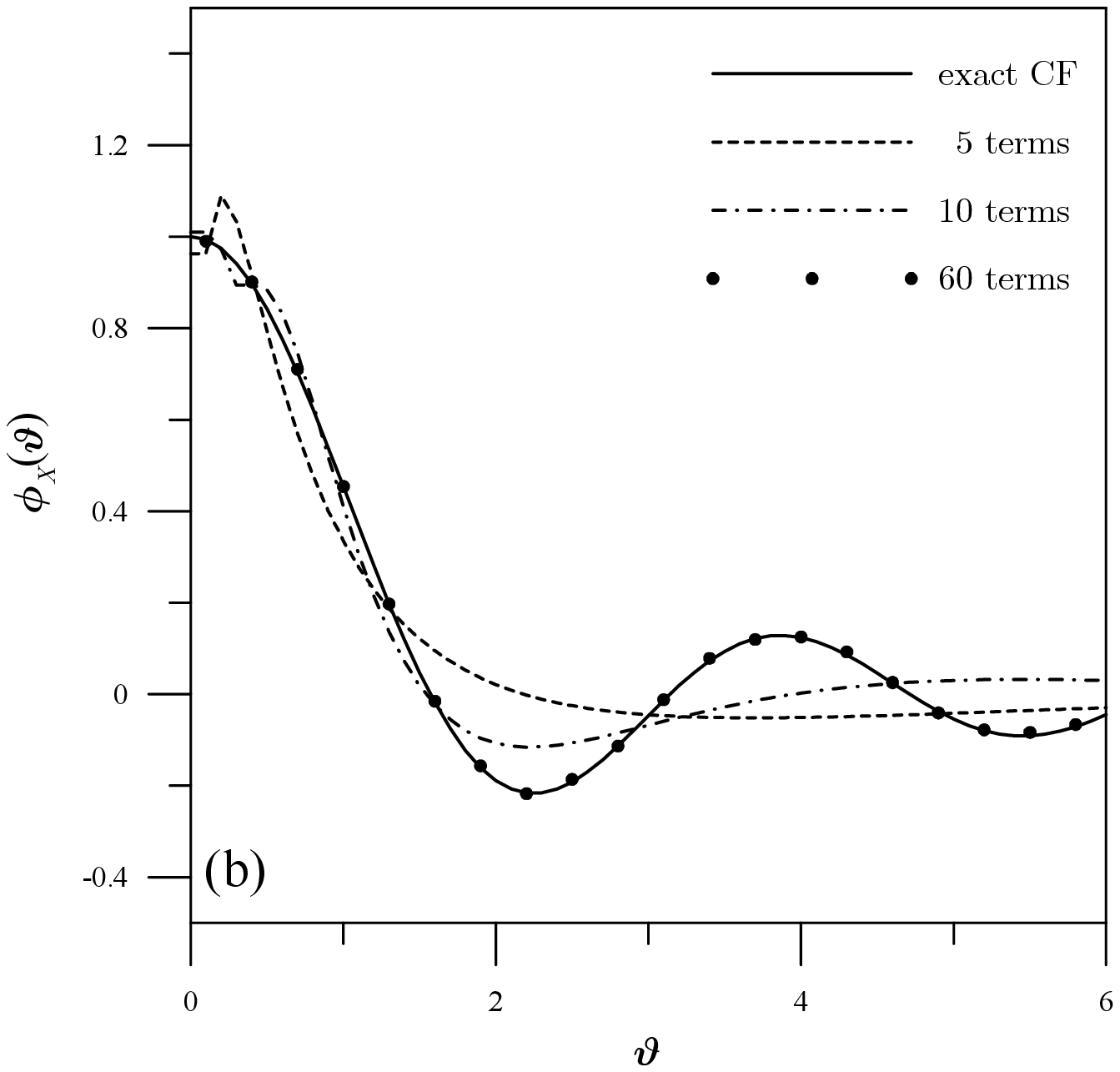}}
  \caption{CF of a uniform distribution in the interval $\left[-2,2\right]$; (3a) Classical Taylor expansion; (3b) Fractional Taylor series (\ref{eq41})--(\ref{eq42}) for different truncation order.} 
 \label{Figure3}
 \end{figure}

\begin{figure}[p]
 \centering
 \subfigure[]
   {\includegraphics[width=6.5cm]{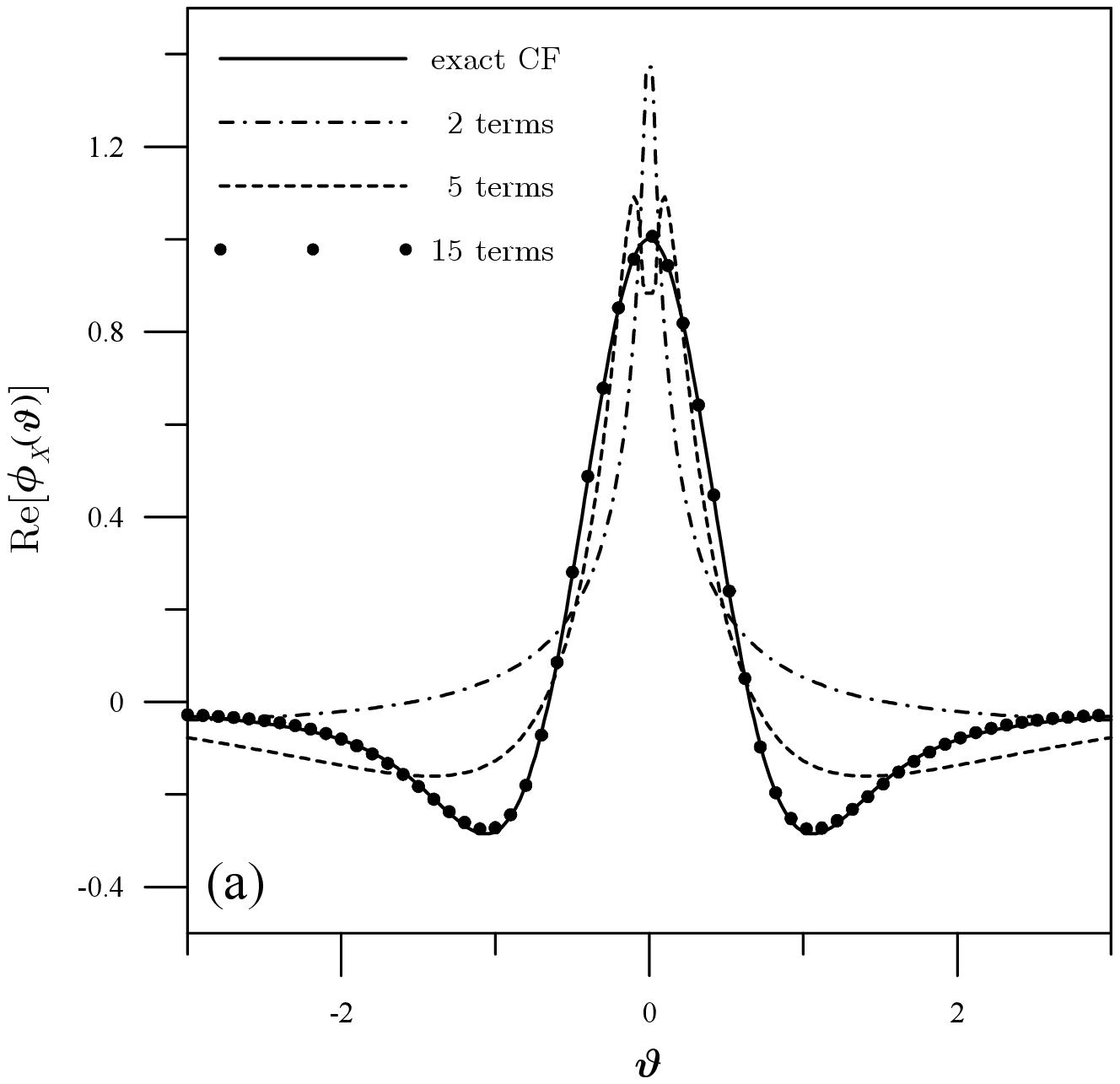}}
 \subfigure[]
   {\includegraphics[width=6.5cm]{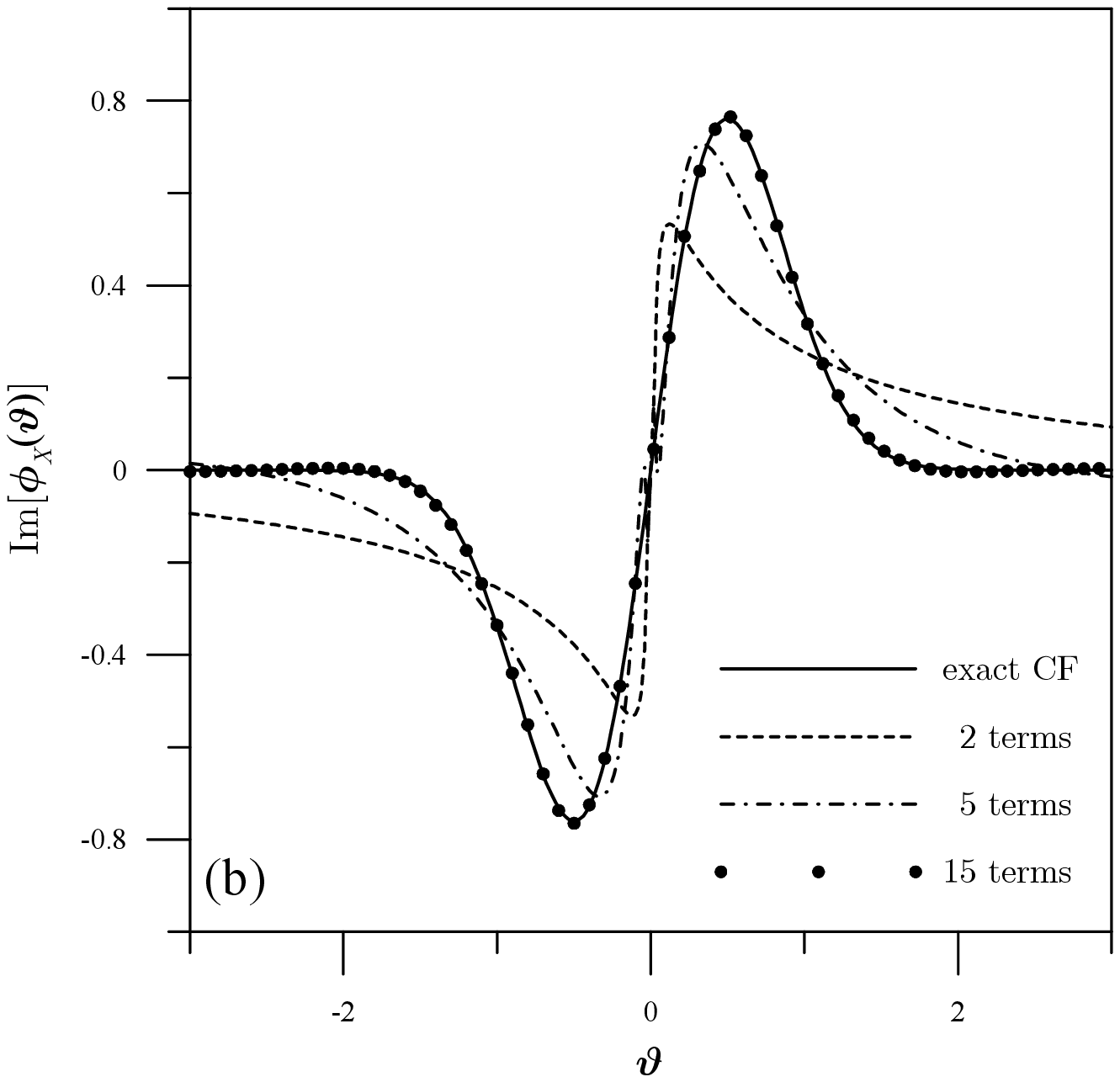}}
  \caption{Fractional Taylor series of the CF of a Rayleigh distribution with scale parameter 2; (4a) Real part; (4b) Imaginary part.} 
 \label{Figure4}
 \end{figure}

\begin{figure}[p]
\centering
\includegraphics[width=8cm]{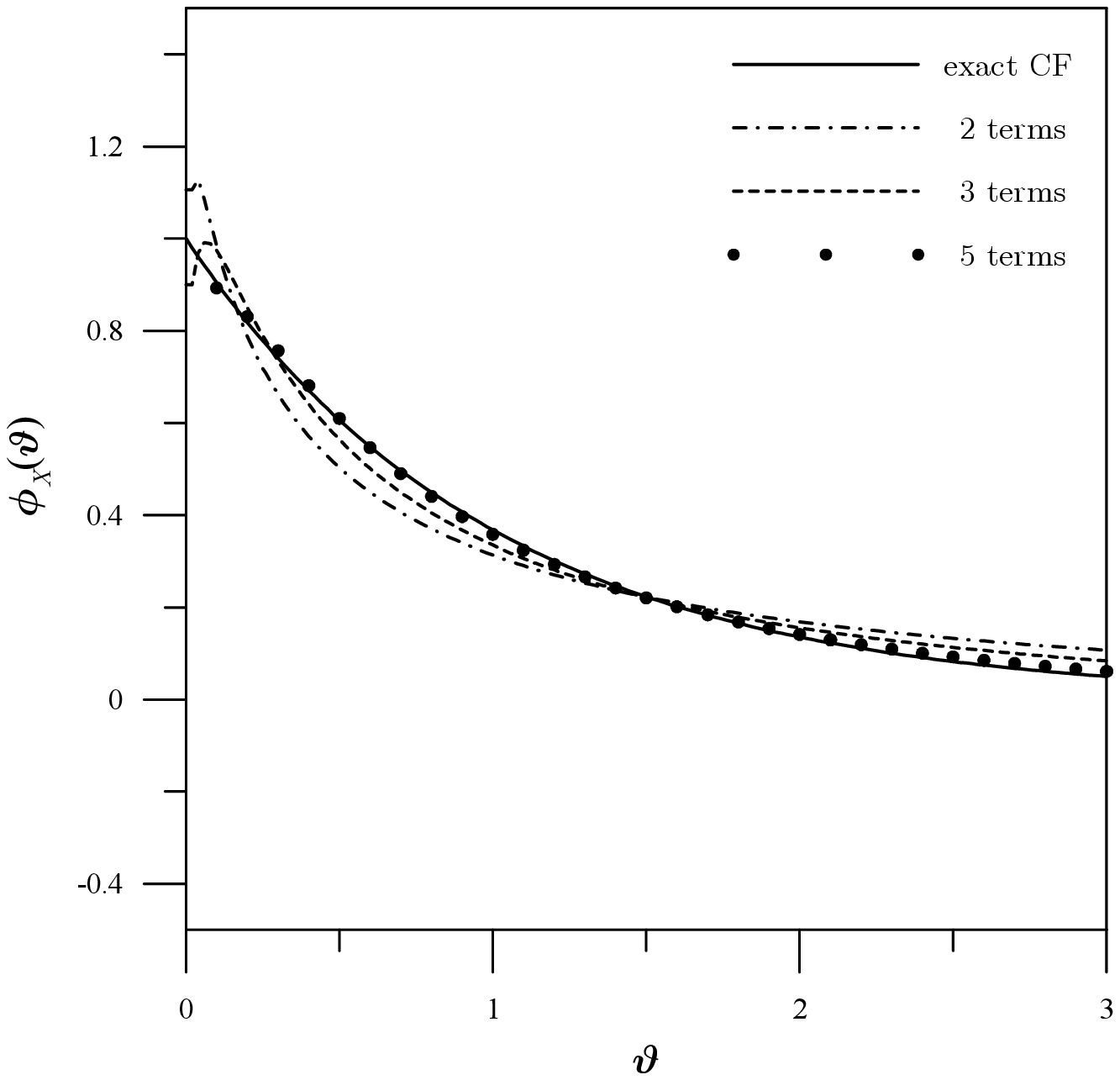}%
\caption{Fractional Taylor series of the CF of a Cauchy distribution with zero location and unitary scale parameter.}%
\label{fig5}%
\end{figure}

\begin{figure}[p]
\centering
 \subfigure[]
   {\includegraphics[width=6.5cm]{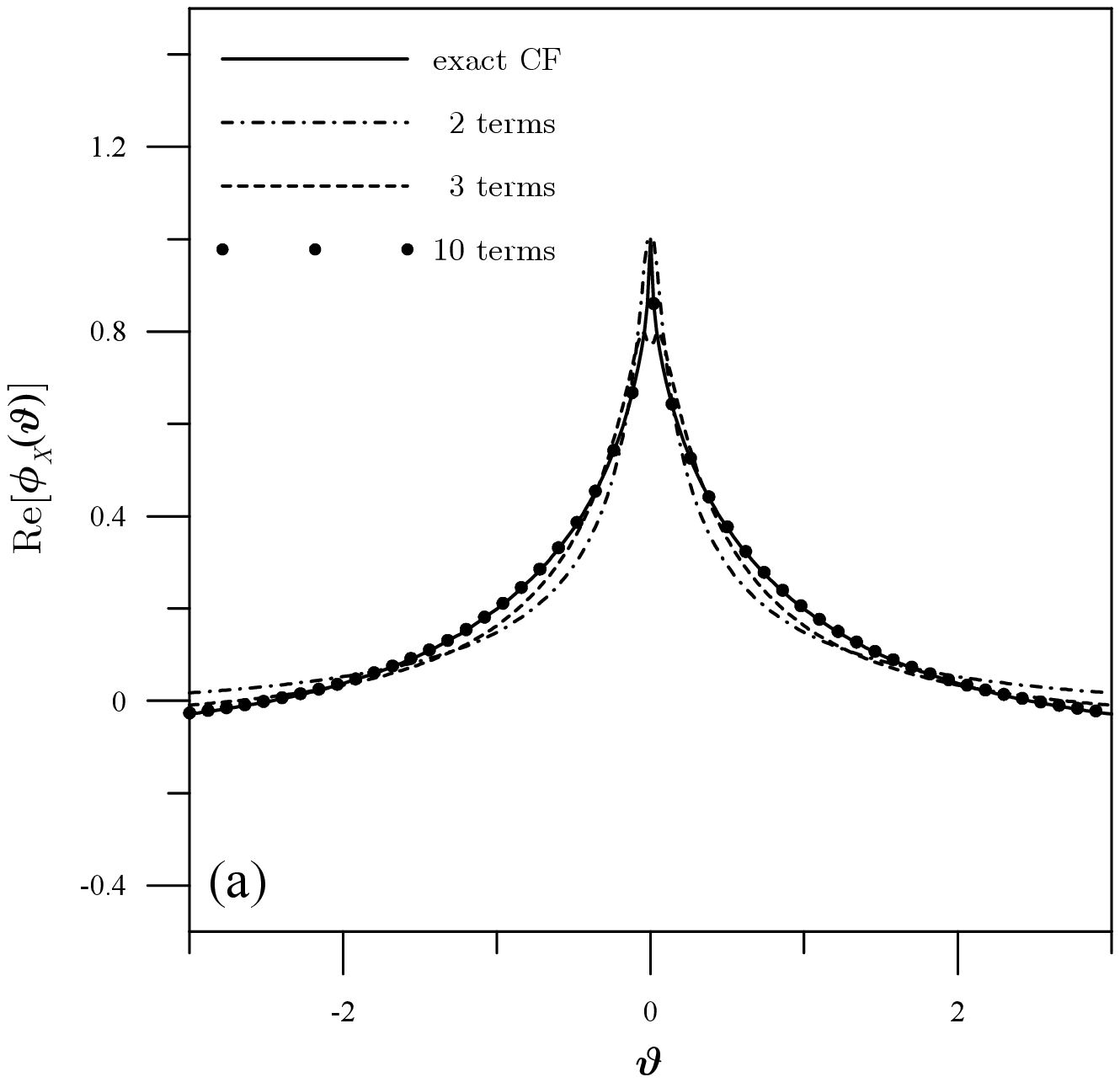}}
 \subfigure[]
   {\includegraphics[width=6.5cm]{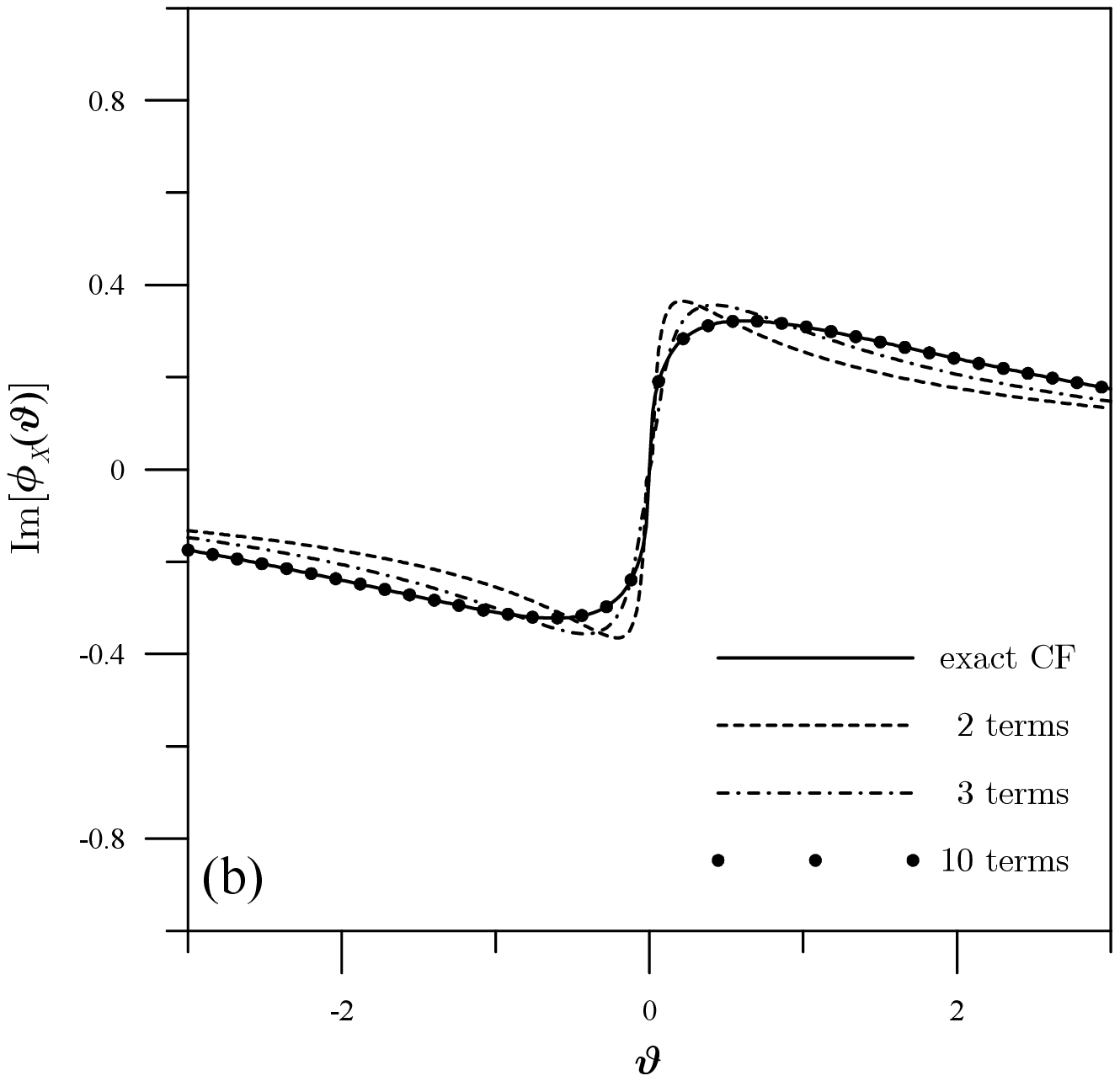}}
  \caption{Fractional Taylor series of the CF of a L\'evy distribution with zero location and unitary scale parameter.} 
 \label{Figure6}
 \end{figure}

\begin{figure}[p]
\centering
\includegraphics[width=8cm]{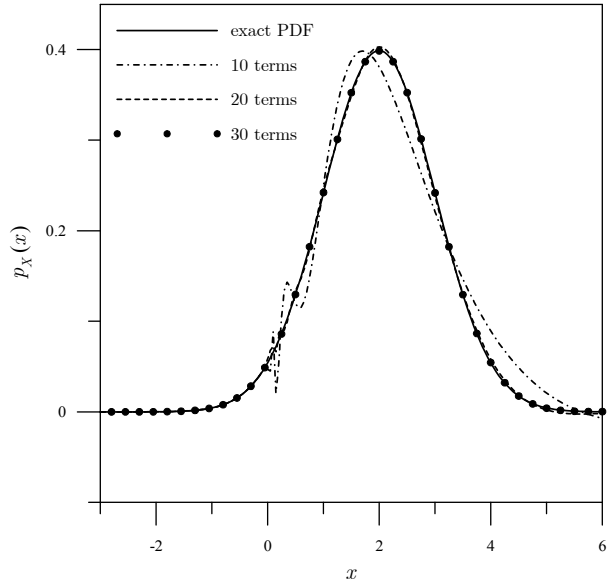}%
\caption{Fractional Taylor series of the PDF of a Gaussian distribution with mean 2 and unitary standard deviation.}%
\label{fig7}%
\end{figure}

\begin{figure}[p]
\centering
\includegraphics[width=8cm]{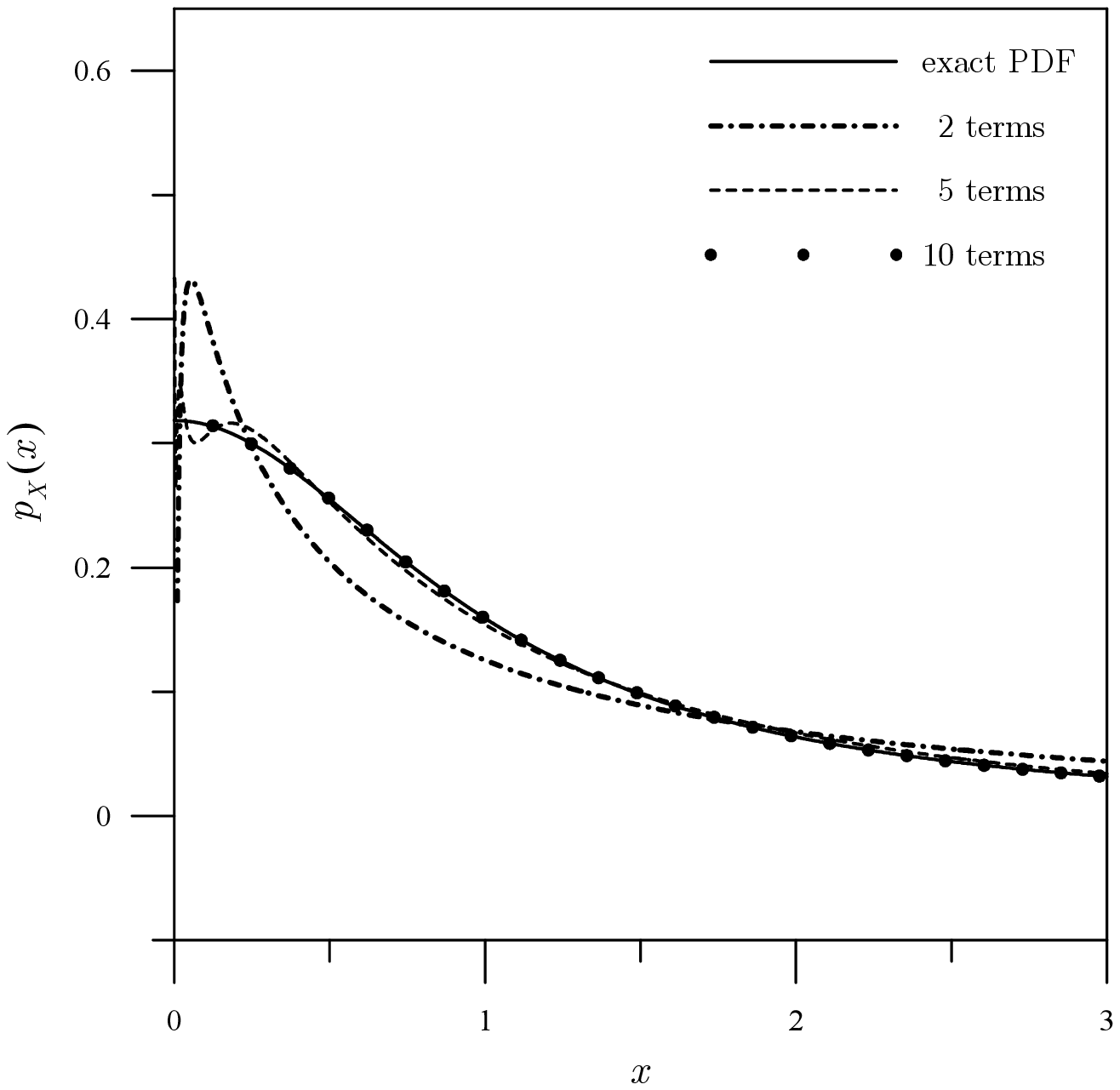}%
\caption{Fractional Taylor series of the PDF of a Cauchy distribution with zero location and unitary scale parameter.
}%
\label{fig8}%
\end{figure}

\begin{figure}[p]
\centering
\includegraphics[width=8cm]{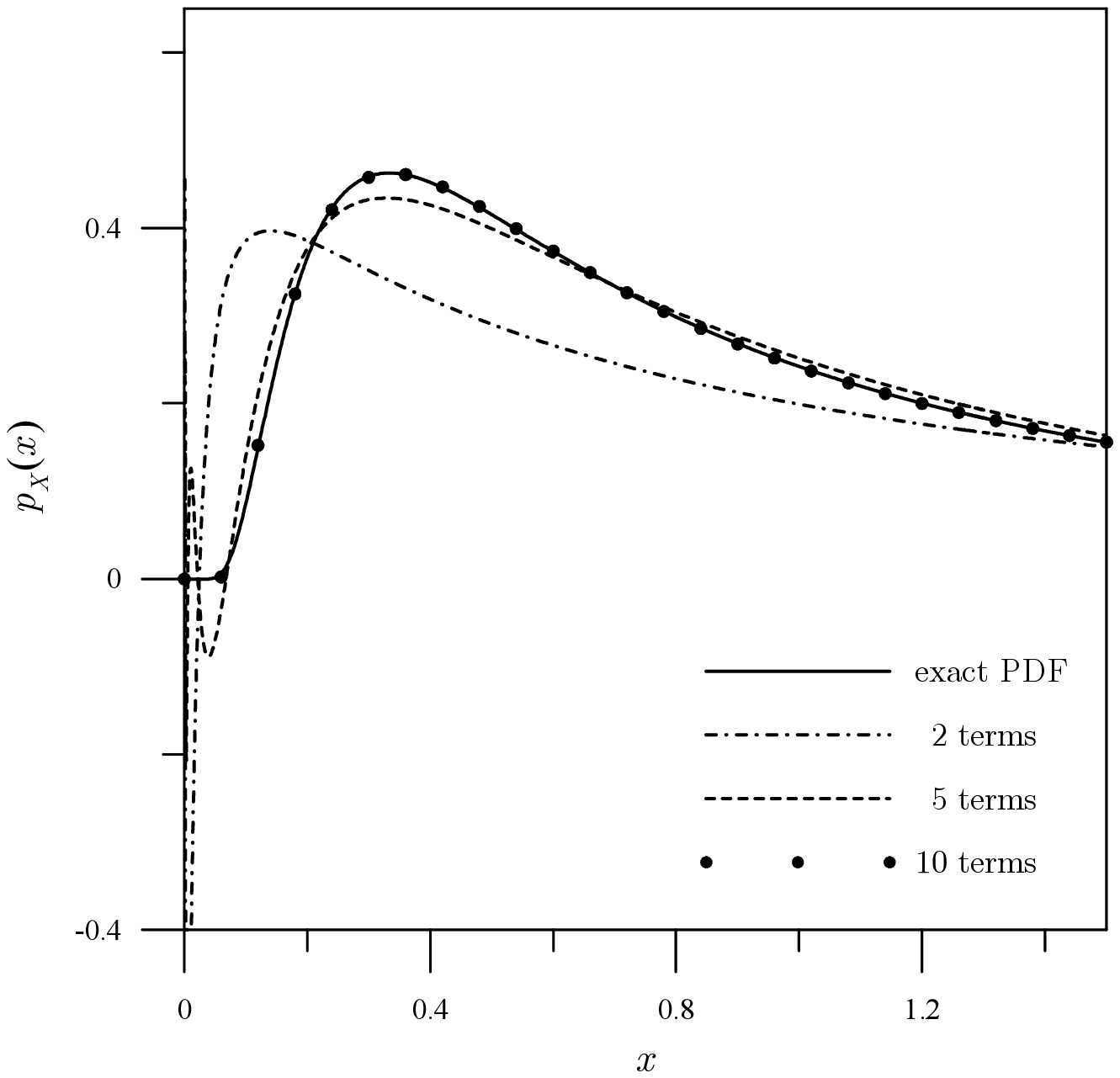}%
\caption{Fractional Taylor series of the PDF of a L\'evy distribution with zero location and unitary scale parameter.}%
\label{fig9}%
\end{figure}

\end{document}